\documentclass[10pt,a4paper]{article}
\usepackage{jheppub_kim}
\usepackage{pdflscape}
\usepackage{amsmath}
\usepackage{amssymb}
\usepackage{dcolumn}
\usepackage{bm}
\usepackage{color}
\usepackage{epsfig}
\usepackage{amsfonts}
\usepackage{graphicx}
\usepackage{subfigure}
\usepackage{dcolumn}
\usepackage{float}
\usepackage{epstopdf}
\usepackage{booktabs}
\usepackage{adjustbox}
\usepackage{lscape}
\usepackage{pdflscape}
\UseRawInputEncoding

\begin{document}
\title{Some New Types of Well-Behaved Polynomial Redshift Parametrization of Dark Energy Equation of State}

\author[a]{Prabir Rudra}
\author[b]{Aritra Sanyal}
\author[c]{Promila Biswas,}
\author[d]{Tuhina Ghorui,}
\author[e]{Ritabrata Biswas,}
\author[f]{Farook Rahaman}

\affiliation[a] {Department of Mathematics, Asutosh College, Kolkata-700 026, India.}

\affiliation[b,c,d,f]{Department of Mathematics, Jadavpur University, Kolkata-700 032, India.}

\affiliation[e]{Department of Mathematics, The University of Burdwan, Golapbag Academic Complex, Burdwan -713104, Purba Barddhaman, West Bengal, India.}

\emailAdd{prudra.math@gmail.com}
\emailAdd{aritrasanyal1@gmail.com}
\emailAdd{promilabiswas8@gmail.com}
\emailAdd{tuhinaghorui.math@gmail.com}
\emailAdd{biswas.ritabrata@gmail.com}
\emailAdd{farookrahaman@gmail.com}

\abstract{In this paper, we explore a new type of smooth and well-behaved polynomial redshift function that can avoid a future singularity. Using this function, we have proposed different redshift parametrizations of the dark energy equation of state, drawing motivation from different polynomial functions like conventional polynomial, Legendre polynomial, Laguerre polynomial, Chebyshev polynomial and Fibonacci polynomial. The main feature of these parametrizations is their well-behaved nature throughout the evolution of the universe, which was a matter of concern in most of the previous polynomial parametrizations of the dark energy equation of state (EoS). This form of parametrization may be considered as an extension of those forms with no divergence at any redshift value. A comprehensive observational data analysis is performed with the Hubble, BAO and DESI datasets to constrain the parameter space of the models. Confidence contours showing joint and marginalized posterior distribution with different combinations of datasets are generated using a Markov Chain Monte Carlo approach. We see that our improved parametrizations enable us to derive more stringent restrictions on the current dark energy EoS and its derivative, which improves performance. Finally, a machine learning analysis is performed using some suitable algorithms like ELR, PILR, ANN, SVR, ERFR and GBR to compare the models. Among all the tested polynomial bases, the Legendre basis demonstrated superior performance with the lowest test RMSE and reduced $\chi^{2}$ value under the Modified Differential Evolution theoretical model, indicating exceptional physical accuracy and numerical stability.}



\maketitle

\section{Introduction}
Observations of Type Ia supernovae (SN) \cite{o1, o2, o3, o4} more than ten years ago revealed that the universe is expanding at an accelerating rate.  Several independent observations have confirmed the Universe's current accelerated expansion, including the previously mentioned Type Ia supernovae \cite{o5, o6, o7, o8, o9}, measurements of cluster properties like mass, correlation function, and evolution with redshift of abundance \cite{o10, o11, o12, o13}, optical surveys of large-scale structure \cite{o14, o15, o16}, anisotropies in the cosmic microwave background (CMB) \cite{o17, o18, o19}, cosmic shear measured from weak lensing \cite{o20, o21}, and Lyman $\alpha$ forest absorption \cite{o22, o23}.

The existence of a new component in the universe, known as dark energy (DE), which would account for around 70–75 \% of the mass–energy of the universe and negate the effects of gravitational pull, could explain this accelerated expansion. The remaining 25–20 \% would primarily consist of dark matter (DM), with the remaining 5 \% going toward baryons, radiation, and neutrinos. Despite the fact that dark energy's existence has been amply demonstrated by observation, no theoretical model that provides a theoretical physical framework for understanding DE, that is consistent with all significant findings, is completely convincing. The simplest hypothesis is the $\Lambda$CDM model, which attributes the acceleration to the existence of a cosmological constant $\Lambda$, and agrees fairly well with observational data. Due to its simplicity and consistency with facts, this model has established itself as the accepted model of cosmology; yet, it is still unable to explain certain phenomena, such as the modest value of the cosmological constant, which cannot be explained by any of the known interactions. 

Other scenarios that acknowledge a slightly time-varying dark energy as the agent of cosmic acceleration are therefore suggested. Quintessence \cite{o24}, Chaplygin gas \cite{o25}, modified gravity \cite{o26}, holographic dark energy \cite{o27}, braneworld models \cite{o28}, $f(R)$ theories \cite{o29}, theories with extra dimensions \cite{o30}, and many more have all surfaced along theoretical lines. Unfortunately, none of them has come up with a clear solution to the DE problem. As a result, while theory is now stuck in a maze of alternatives, empirical research is making rapid progress in terms of both the quantity and quality of data. Since it is unclear exactly what the observations should be compared to, a tried-and-tested method taken from many branches of physics is the definition of a parametrization of DE functions. If done correctly, this method enables one to condense all of the observational information into a few numbers that can then be compared to theoretical predictions.

Examining the DE equation of state (EoS) is the most logical course of action in this situation. It is often believed that this quantity fluctuates slowly with redshift and hence can be roughly represented by a fitting formula with a limited number of free parameters. By employing an optimization process to compare the ansatz with data, these parameters can be constrained. Although choices are usually based on past knowledge and intuition, there is always an opportunity for debate and advancement. This is exactly the path we take in this work.

There are a few well-known DE parametrizations available in the literature with different properties and merits of their own. Some of the major ones are Chevallier-Polarski-Linder (CPL) parametrization \cite{p3, p4}. Here, the DE equation of state is given by
\begin{equation}
 w(z)=w_{0}+w_{1}\frac{z}{1+z}   
\end{equation}
where $w_{0}$ is the current value (at redshift $z=0$) of DE EoS. Due to its simplicity, sensitivity to observational data, good behavior, and boundedness at high redshifts, this parametrization has found widespread use. Its acceptance as a recommended parametrization by the Dark Energy Task Force \cite{detf} has helped to increase its appeal. In addition to the fact that one of the parameters of the CPL parametrization is usually subject to considerable percentage limits, it suffers from a fairly significant correlation. However, it has some good properties; thus, a convenient redefinition was proposed to address these two problems and produce a better outcome, even though the encoded data is still exactly the same. This was accomplished by Wang in \cite{wang}, which provided a novel description of the dark energy in terms of its current value, $w_{0}$ at $z=0$, and $w_{0.5}$ at $z=0.5$. Other than the CPL parametrization, we have the Jassal-Bagla-Padmanabhan (JBP) \cite{p5} where the DE EoS is given by
\begin{equation}
 w(z)=w_{0}+w_{1}\frac{z}{\left(1+z\right)^{2}}   
\end{equation}
When applying the restrictions from WMAP measurements to type Ia Supernovae (SNeIa) observations, it is demonstrated that this model limits the enormous range of energy density at low redshift. Considerable differences between the CPL and JBP parametrizations occur at high redshift. We see that $w(\infty)=w_{0}+w_{1}$ for CPL, but for JBP $w(\infty)=w_{0}$. As a result, JBP parametrization can simulate a dark energy component that exhibits rapid variation at low $z$ but has the same equation of state at high redshifts at the current epoch. In \cite{p7} the Barboza and Alcaniz have introduced a DE parametrization given by
\begin{equation}
w(z)=w_{0}+w_{1}\frac{z\left(1+z\right)}{1+z^{2}}    
\end{equation}
For this model, the universe undergoes a transition from deceleration to acceleration at redshift $z\approx 0.58$. Apart from this, there are other DE EoS parametrizations like Efstathiou Redshift Parameterization (ERP) or Log Parametrizations \cite{efs}, Alam-Sahni-Saini-Starobinsky (ASSS) Redshift Parameterization \cite{asss, asss2}, etc.

In \cite{p6}, the authors have introduced two different polynomial parametrizations of DE EoS with two free parameters and explored their correlation properties. The free parameters were constrained using observational data, and it was shown that the new polynomial parametrizations are considerable improvements on the well-known CPL parametrization. Moreover, it was also shown that the polynomial models performed well over the CPL re-parametrization proposed by Wang. But due to the presence of the terms $\frac{1}{1+z}$ these EoS models become singular in the future epoch at $z=-1$. This can be a serious pathology while analyzing the evolution properties of these models, especially in the future universe. Motivated by this, here we would like to discuss a new type of well-behaved polynomial redshift function, which can safely avoid this future singularity. Then, using the ansatz, we will develop some new polynomial parametrization models having their roots in deep mathematical analysis. We will constrain the parameter space of these models using observational data and perform a comparative study using machine learning algorithms. The paper is organized as follows: In Section 2, we discuss the gravitational field equations and the corresponding cosmological parameters that we will use in our study. In Section 3, we propose a new well-behaved polynomial redshift function. Using this polynomial function, we propose some redshift EoS for DE. A detailed observational data analysis is performed in Section 4 using observational data to constrain the parameter space of the models. In Section 5, we will apply machine learning algorithms to reconstruct the Hubble function and compare their performance with our theoretical models. Finally, the paper ends with some concluding remarks in section 6.

\section{Field Equations}
In the context of General Relativity (GR), the Friedmann-Robertson-Walker (FRW) equations explain the expansion of a homogeneous and isotropic universe. The FRW equations are given by 
\begin{equation}
H^{2}=\frac{8\pi G}{3}\rho-\frac{k}{a^{2}}
\end{equation}
and
\begin{equation}
\frac{\ddot{a}}{a}=-\frac{4\pi G}{3}\left(\rho+3p\right)
\end{equation}
where $H=\frac{\dot{a}}{a}$ is the Hubble parameter and $a(t)$ is the cosmological scale factor. $G$ is the Newton's gravitation constant and $k$ determines the curvature of the universe. For $k=0,1,-1$ we have respectively the flat, closed and open universe. Moreover, $\rho$ and $p$ denote respectively the energy density and pressure of matter. Presently the main matter components are dark matter (DM) and dark energy (DE). The late cosmic acceleration is attributed to the DE component of matter and hence is the most important component under study. So the total energy density and pressure can be given by $\rho=\rho_{de}+\rho_{m}$ and $p=p_{de}+p_{m}$. Since the pressure of matter is considered to be negligible now, i.e., $p_{m}\approx0$, we have $p=p_{de}$.

The behavior of DE is determined by its equation of state (EOS) given by \cite{p1, p2}
\begin{equation}
w(z)=\frac{p_{de}}{\rho_{de}}
\end{equation}
where $z$ is the cosmological redshift parameter given by 
\begin{equation}\label{redshift}
z=\frac{1}{a(t)}-1\end{equation}. The EOS parameter determines the form of the Hubble parameter $H(z)$ or any of its forms necessary to obtain observable quantities. For a flat universe we have
\begin{equation}\label{hubb}
\frac{H^{2}(z)}{H_{0}^{2}}=\Omega_{m}\left(1+z\right)^{3}+\Omega_{de}X(z)
\end{equation}
where $X(z)=\frac{\rho_{de}(z)}{\rho_{de}(0)}=exp\left(3\int_{0}^{z}\frac{1+w(z)}{1+z}dz\right)$
and $\Omega_{m}+\Omega_{de}=1$. Here $\Omega_{i}$ is a dimensionless density parameter given by $\Omega_{i}=\frac{\rho_{i}}{\rho_{c}}$ with $\rho_{c}=3H_{0}^{2}/8\pi G$. $H_{0}$ is the present value of the Hubble parameter.

The conservation equation for each component is given by,
\begin{equation}
\frac{\dot{\rho}}{\rho}=-3H\left(1+\frac{p}{\rho}\right)=-3H\left(1+w(z)\right)
\end{equation}

The deceleration parameter can be given by
\begin{equation}
q(z)=-\frac{\ddot{a}}{aH^{2}}=\frac{1}{2}\left[1+3w(z)\Omega_{de}\right]
\end{equation}
In the subsequent section, we will propose a new smooth and well-behaved polynomial redshift function that avoids any future singularity and then use it to propose some parametrized redshift DE EoS. Later, a detailed observational data analysis will be performed to study the models.

\section{A new Polynomial Parametrization}
The most common form of this type of parametrization is
\begin{equation}
w(z)=-1+c_{1}\left(1+z\right)+c_{2}\left(1+z\right)^{2}
\end{equation}
Introducing some improvements a generalized form becomes
\begin{equation}
w(z)=-1+c_{1}\left(1+f(z)\right)+c_{2}\left(1+f(z)\right)^{2}    
\end{equation}
A further generalized form will take the form
\begin{equation}
w(z)=-1+c_{1}g_{1}\left(1+f(z)\right)+c_{2}g_{2}\left(1+f(z)\right)
\end{equation}
where $g_{1}$ and $g_{2}$ are smooth functions. We think that it is reasonable and convenient to confine (at least initially) the parameter search region to $\mid c_{1} \mid <1$ and $\mid c_{2} \mid <1$ because one does not expect a large departure from the $\Lambda$CDM setting.

Using our idea of well-behaved functions, we propose
\begin{equation}
f(z)=\frac{z}{1+z^{2}}
\end{equation}
By this, we avoid the high redshift unboundedness while at the same time keeping the function well-behaved throughout the evolution of the universe, i.e. $z ~\epsilon~ [-1,\infty)$.

\subsection{Conventional Polynomial Parametrization}
Using our ansatz, we have the following parametrization
\begin{equation}
w(z)=-1+c_{1}\left(1+\frac{z}{1+z^{2}}\right)+c_{2}\left(1+\frac{z}{1+z^{2}}\right)^{2}    
\end{equation}
The above parametrization can be written in a compact form as
\begin{equation}\label{conventional}
w(z)=-1+c_{1}\left(\frac{1+z+z^{2}}{1+z^{2}}\right)+c_{2}\left(\frac{1+z+z^{2}}{1+z^{2}}\right)^{2}    
\end{equation}
We can see that for $c_{1}=c_{2}=0$ the above form reduces to $\Lambda$CDM model. In fact, the first term is taken as $-1$ to facilitate this transition and help us compare the models with the standard ones. Thus, this form is better than the standard parametrizations found in the literature. This form removes the unbounded nature of linear parametrization at high redshifts. Moreover, being well-behaved throughout the domain of evolution of the universe, i.e., $z ~\epsilon~ [-1,\infty)$, the model scores above the well-known parametrizations like the CPL \cite{p3, p4} and JBP \cite{p5} models. It should be stated here that in selecting this model, we have been inspired by the parametrizations of Feng-Shen-Li-Li (FSLL) \cite{p6} and Barboza-Alcaniz (BA) \cite{p7}. In fact, we have extended the polynomial parametrization using the FSLL and BA parametrizations to improve the setting. Moreover, by doing this, we have kept the generality of the polynomial parametrization intact.

For the present universe, ($z=0$) we see that for our model $w(0)=-1+c_{1}+c_{2}$. Moreover for high redshifts as $z\rightarrow \infty$, working out the limit we get $w(\infty)\rightarrow -1+c_{1}+c_{2}$. This shows that early-time inflation and late-time acceleration are somewhat similar types of events driven by some exotic matter of nearly identical nature. We also get $w'(0)=c_{1}+2c_{2}$, and for $z\rightarrow \infty$ we get $w'(\infty)\rightarrow 0$, which shows the rate at which the EoS changes in the present time and in the extreme past. Here we get a difference in the two scenarios. In the present time, there is a substantial rate at which the dark energy EoS changes, whereas in the early times, the EoS was nearly stagnant for a considerable period, showing almost a constant EoS of dark energy.

\subsection{Legendre Polynomial Parametrization}
Legendre polynomials are a system of complete and orthogonal polynomials with many mathematical properties and a variety of uses in mathematics. There are numerous ways to characterize them, and each explanation emphasizes a distinct facet while also pointing out linkages to various mathematical structures, physical applications, and numerical applications.
Here, using the Legendre polynomial, we propose a parametrization of the form
\begin{equation}
w(z)=-1+c_{1}P_{1}(1+f(z))+c_{2}P_{2}(1+f(z))    
\end{equation}
where $P_{n}$ are the Legendre polynomial of first kind with degree $n$. Simplifying the above, we get
\begin{equation}\label{legendre}
w(z)=-1+c_{1}\left(\frac{1+z+z^{2}}{1+z^{2}}\right)+\frac{c_{2}}{2}\left[3\left(\frac{1+z+z^{2}}{1+z^{2}}\right)^{2}-1\right]
\end{equation}
where we have used $P_{1}(x)=x$ and $P_{2}(x)=\frac{1}{2}\left(3x^{2}-1\right)$.
Like the previous model, we can see that for $c_{1}=c_{2}=0$ the above form reduces to $\Lambda$CDM model.

For the present universe ($z=0$) we see that for our model $w(0)=-1+c_{1}+c_{2}$. Moreover for high redshifts as $z\rightarrow \infty$, we get $w(\infty)\rightarrow -1+c_{1}+c_{2}$. The limiting values are similar to the previous case. We also get $w'(0)=c_{1}+3c_{2}$, and for $z\rightarrow \infty$ we get $w'(\infty)\rightarrow 0$, which shows the rate at which the EoS changes in the present time and in the extreme past. We see that the limiting conditions are similar to the previous model.

\subsection{Laguerre Polynomial Parametrization}
In mathematics, the Laguerre polynomials are nontrivial solutions of Laguerre's differential equation, which is a second-order linear differential equation. In quantum mechanics, the Laguerre polynomials appear in the radial portion of the Schrödinger equation solution for an atom with one electron. In phase space, they also explain the static Wigner functions of oscillator systems in quantum mechanics. Here we use Laguerre polynomials and propose the following form 
\begin{equation}
w(z)=-1+c_{1}L_{1}(1+f(z))+c_{2}L_{2}(1+f(z))    
\end{equation}
where $L_{n}$ are the Laguerre polynomial with degree $n$. Simplifying the above we get
\begin{equation}\label{laguerre}
w(z)=-1-c_{1}\left(\frac{z}{1+z^{2}}\right)+\frac{c_{2}}{2}\left[\left(\frac{z}{1+z^{2}}\right)^{2}-2\left(\frac{z}{1+z^{2}}\right)-1\right]
\end{equation}
where we have used $L_{1}(x)=1-x$ and $L_{2}(x)=\frac{1}{2}\left(x^{2}-4x+2\right)$. Here also for $c_{1}=c_{2}=0$ the above form reduces to $\Lambda$CDM model.

For the present universe ($z=0$) we see that for our model $w(0)=-1-\frac{c_{2}}{2}$. Moreover for high redshifts as $z\rightarrow \infty$, we get $w(\infty)\rightarrow -1-\frac{c_{2}}{2}$. The trend of the limiting values are similar to the previous cases. We also get $w'(0)=-c_{1}-c_{2}$, and for $z\rightarrow \infty$ we get $w'(\infty)\rightarrow 0$, which shows the rate at which the EoS changes in the present time and in the extreme past. We see that the limiting conditions are of a similar nature to the previous model.

\subsection{Chebyshev Polynomial Parametrization}
Chebyshev polynomials are useful tools in fields like machine learning, control theory, signal processing, and optimization that call for effective numerical techniques and approximations.  They are useful for real-world computer issues because of their effectiveness in reducing errors and generating reliable approximations.

\begin{equation}
w(z)=-1+c_{1}T_{1}(1+f(z))+c_{2}T_{2}(1+f(z))    
\end{equation}

\begin{equation}\label{chebyshev}
w(z)=-1+c_{1}\left(\frac{1+z+z^{2}}{1+z^{2}}\right)+c_{2}\left[2\left(\frac{1+z+z^{2}}{1+z^{2}}\right)^{2}-1\right]
\end{equation}
where we have used $T_{1}(x)=x$ and $T_{2}(x)=2x^{2}-1$. For $c_{1}=c_{2}=0$ the above form reduces to the $\Lambda$CDM model.
For the present universe ($z=0$) we see that for our model $w(0)=-1+c_{1}+c_{2}$. Moreover for high redshifts as $z\rightarrow \infty$, we get $w(\infty)\rightarrow -1+c_{1}+c_{2}$. We also get $w'(0)=c_{1}+4c_{2}$, and for $z\rightarrow \infty$ we get $w'(\infty)\rightarrow 0$, which shows the rate at which the EoS changes in the present time and in the extreme past.

\subsection{Fibonacci Polynomial Parametrization}
Like the Fibonacci numbers themselves, Fibonacci polynomials are a series of polynomials, each of which is defined in terms of the two polynomials that came before it in the sequence. These polynomials have a number of interesting and useful applications in various fields like signal processing, numerical methods, combinatorics, mathematical physics, control theory and robotics, coding theory, cryptography, etc. The main attraction of the Fibonacci polynomials is their recursive nature and their applicability to a wide variety of issues in different domains that deal with growth patterns, recursion, or polynomials.
\begin{equation}
w(z)=-1+c_{1}F_{1}(1+f(z))+c_{2}F_{2}(1+f(z))    
\end{equation}

\begin{equation}\label{fibonacci}
w(z)=-1+c_{1}+c_{2}\left(\frac{1+z+z^{2}}{1+z^{2}}\right)
\end{equation}
where we have used $F_{1}(x)=1$ and $F_{2}(x)=x$. For $c_{1}=c_{2}=0$ the above form reduces to the $\Lambda$CDM model.
For the present universe ($z=0$) we see that for our model $w(0)=-1+c_{1}+c_{2}$. Moreover for high redshifts as $z\rightarrow \infty$, we get $w(\infty)\rightarrow -1+c_{1}+c_{2}$. We also get $w'(0)=c_{2}$, and for $z\rightarrow \infty$ we get $w'(\infty)\rightarrow 0$, which shows the rate at which the EoS changes in the present time and in the extreme past.

\section{Observational Data Analysis}
In this section, we will perform an observational data analysis with different data sets to constrain our cosmological models. We will use $\chi^{2}$ minimization technique along with the observational to put constraints on the parameter space of the models.The model parameters are constrained using a Markov Chain Monte Carlo (MCMC) approach implemented via the \texttt{emcee} sampler \cite{Foreman-Mackey:2012any}. The posterior distributions are visualized using the \texttt{GetDist} package \cite{Lewis:2019xzd}. For each dataset and their combinations, we compute the best-fit values and corresponding  confidence intervals, summarized in Table.

The likelihood function is defined as
\begin{equation}
    \mathcal{L} \propto \exp\left(-\frac{1}{2}\chi^2\right),
\end{equation}
where the chi-square function is given by
\begin{equation}
    \chi^2 = \Delta \mathbf{D}^T \mathbf{C}^{-1} \Delta \mathbf{D},
\end{equation}
with \(\Delta \mathbf{D} = \mathbf{D}_{\text{obs}} - \mathbf{D}_{\text{th}}\), and \(\mathbf{C}\) denoting the covariance matrix for each dataset.

\subsection{Datasets and Confidence Contours}
For the observational analysis, we employed three independent and complementary cosmological datasets that are widely used for constraining the Hubble parameter $H(z)$ and dark energy equation of state (EoS) models. The combined use of these datasets allows for a robust estimation of the model parameters with minimized degeneracies.

To constrain the free parameters of the proposed polynomial dark energy models, we employ three complementary cosmological datasets: the Hubble parameter ($H(z)$) measurements, the Baryon Acoustic Oscillation (BAO) data, and the recent Dark Energy Spectroscopic Instrument (DESI) measurements. These datasets probe different cosmic epochs and physical scales, thereby ensuring statistically robust and cross-verified constraints on the cosmological parameters.

\begin{itemize}
    \item \textbf{HUBBLE Data:} We use 32 model-independent measurements of the Hubble parameter \( H(z) \), commonly known as Cosmic Chronometers (CC) \cite{Jimenez2002, Moresco2015}. The covariance matrix used to estimate the likelihood is constructed following the approach described in \cite{Moresco2012, Moresco2016}. These data points provide direct constraints on the expansion history of the universe.
    
    \item \textbf{BAO Data:} The BAO dataset provides constraints on the large-scale structure of the universe through measurements of the baryon acoustic feature imprinted in the matter power spectrum. We use the distance measurements from various galaxy surveys compiled in the DESI and eBOSS analyses \cite{DESI2024, eBOSS2021}. The observables considered include \( D_M/r_d \), \( D_H/r_d \), and \( D_V/r_d \), where \( D_M \) is the comoving angular diameter distance, \( D_H \) the Hubble distance, and \( r_d \) the comoving sound horizon at the drag epoch.
    
    \item \textbf{DESI Data:} The Dark Energy Spectroscopic Instrument (DESI) Release II dataset provides high-precision BAO and redshift-space distortion (RSD) measurements across a wide redshift range \cite{DESI2025}. These measurements significantly improve constraints on late-time cosmic acceleration and expansion rate, complementing the HUBBLE and BAO datasets.
\end{itemize}

For parameter estimation, we consider several combinations of observational datasets to assess the robustness of each polynomial basis. The individual datasets include Hubble, BAO, and DESI measurements. We also analyze their combined forms---Hubble+BAO, Hubble+DESI, BAO+DESI, and the joint dataset Hubble+BAO+DESI---to provide a comprehensive evaluation of the model’s performance across different observational constraints. Each dataset combination provides independent constraints on the parameters $\{c_i, H_0\}$, and the final combined analysis (HUBBLE+BAO+DESI) yields the tightest confidence bounds with the smallest statistical uncertainties.
The two-dimensional confidence contours shown in the parameter constraint plots represent the joint probability distributions of the model parameters (e.g., $c_1$--$c_2$, $c_3$--$c_4$, etc.) obtained from the likelihood analysis. The colour scheme of the contour plots follows the standard cosmological convention:
\begin{itemize}
    \item The light-shaded region corresponds to the 95\% confidence level (C.L.), indicating the parameter space within which the true values are expected to lie with 95\% probability.
    \item The dark-shaded region corresponds to the 68\% confidence level (C.L.), representing the $1\sigma$ range of the best-fit values.
\end{itemize}

The central black dot in each contour marks the best-fit parameter values, while the surrounding elliptical regions illustrate the covariance between the two fitted parameters. The darker inner contour thus represents the most probable region of parameter space consistent with the observational data, while the lighter outer contour shows the extended range allowed by the data at a higher uncertainty level. All contour plots in this analysis ensure statistical consistency and reproducibility. The colours were chosen to maintain clarity between the 68\% and 95\% confidence regions across all polynomial bases. Next, we will analyze each model separately and put constraints on the parameter spaces.

\subsection{Conventional Polynomial Parametrization}
Using eqns.(\ref{hubb}) and (\ref{conventional}), we write the dimensionless Hubble parameter for this model as
\begin{equation}
\frac{H^{2}(z)}{H_{0}^{2}}=\Omega_{m}\left(1+z\right)^{3}+\Omega_{de}\left(1+z\right)^{3/4\left(2c_{1}+c_{2}\right)}\left(1+z^{2}\right)^{3/8\left(2c_{1}+3c_{2}\right)}exp\left[3/4\left(\frac{c_{2}z\left(z-1\right)}{1+z^{2}}+2\left(c_{1}+2c_{2}\right) tan^{-1}(z)\right)\right]~~.
\end{equation}
At zero redshift, we obtain the constraint $\Omega_{m0}+\Omega_{de0}=1$. In Table 1, we have presented the best fit values of the free parameters $c_{1}$ and $c_{2}$ for this model, along with the constrained values of the Hubble parameter $H_{0}$ for different combinations of datasets. In Fig.(1) we have presented the confidence contours showing the joint and marginalized posterior distribution with different datasets. The light-shaded region corresponds to the 95 \% confidence level, while the dark-shaded region corresponds to the 68 \% confidence level. In Fig.(2) similar confidence contours are presented for different combinations of the datasets taken two at a time. Finally, in Fig.(3) confidence contours are given for combined Hubble+BAO+DESI datasets.

\begin{table}[ht]
\centering
\caption{Parameter Constraints Summary  for Conventional Polynomial Parametrization}
\begin{tabular}{lccc}
\hline
\hline
Dataset & $c_1$ & $c_2$ & $H_0$ [km/s/Mpc] \\
\hline
HUBBLE            & $0.0209 \pm 0.0789$ & $0.0119 \pm 0.0635$ & $69.71 \pm 1.28$ \\
BAO               & $0.0121 \pm 0.0660$ & $0.0191 \pm 0.0502$ & $69.14 \pm 1.15$ \\
DESI              & $0.0174 \pm 0.0752$ & $0.0142 \pm 0.0534$ & $69.17 \pm 1.23$ \\
HUBBLE+BAO        & $0.0157 \pm 0.0594$ & $0.0133 \pm 0.0469$ & $69.34 \pm 1.08$ \\
HUBBLE+DESI       & $0.0188 \pm 0.0632$ & $0.0102 \pm 0.0504$ & $69.38 \pm 1.14$ \\
BAO+DESI          & $0.0116 \pm 0.0542$ & $0.0192 \pm 0.0402$ & $69.08 \pm 1.01$ \\
HUBBLE+BAO+DESI   & $0.0154 \pm 0.0494$ & $0.0140 \pm 0.0364$ & $69.22 \pm 0.91$ \\
\hline
\multicolumn{4}{l}{\footnotesize Note: $\Omega_m = 0.31$ and $\Omega_{de} = 0.69$ are fixed parameters in this analysis.}
\end{tabular}
\label{tab:parameter_constraints}
\end{table}

\begin{figure}[h!]
    \centering
    \includegraphics[width=\linewidth]{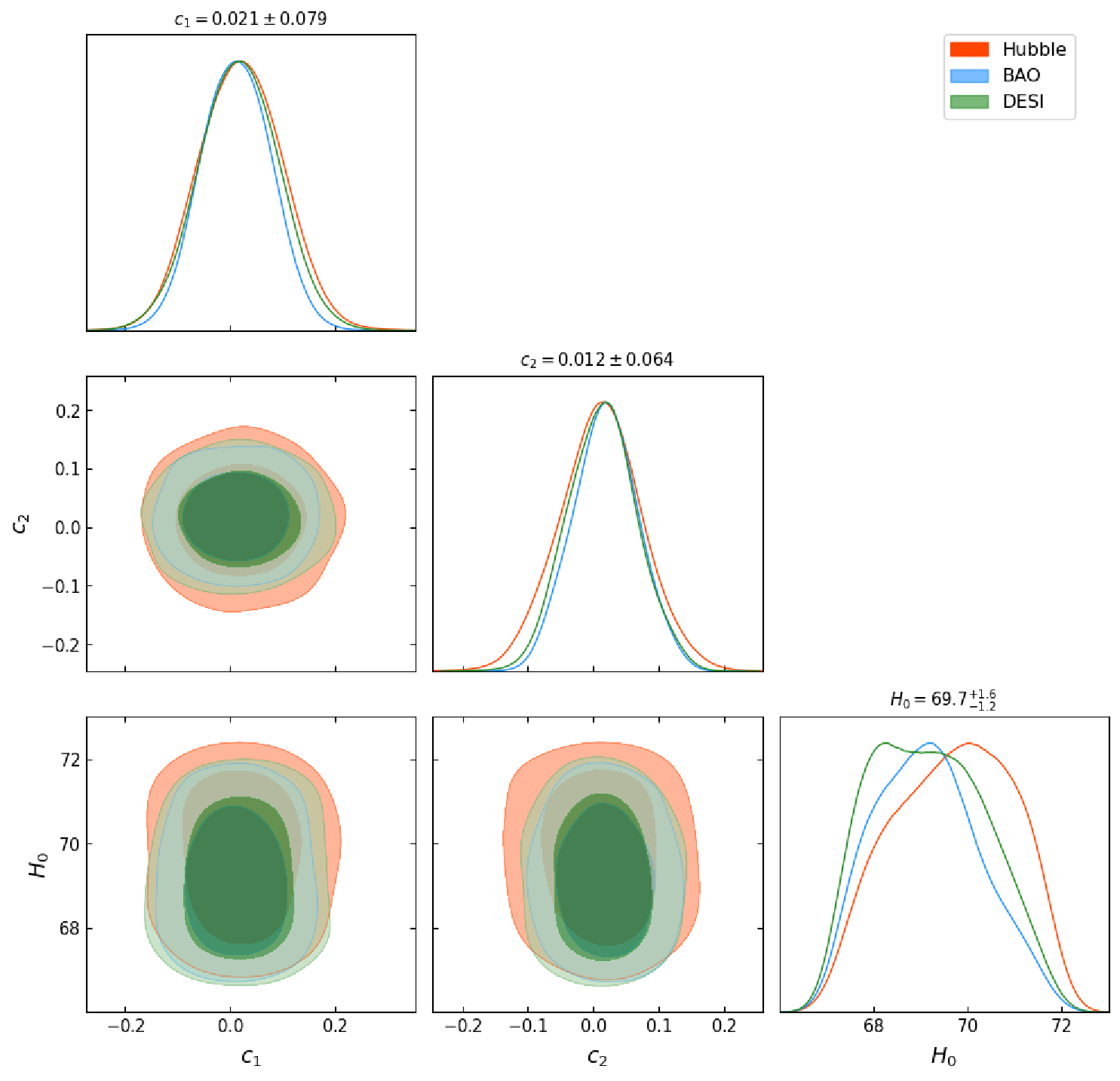}
    \caption{Joint and marginalized posterior distributions  with combinations of datasets Hubble, BAO and DESI for Conventional Polynomial Parametrization}
    \label{fig:placeholder}
\end{figure}

\begin{figure}[h!]
    \centering
    \includegraphics[width=\linewidth]{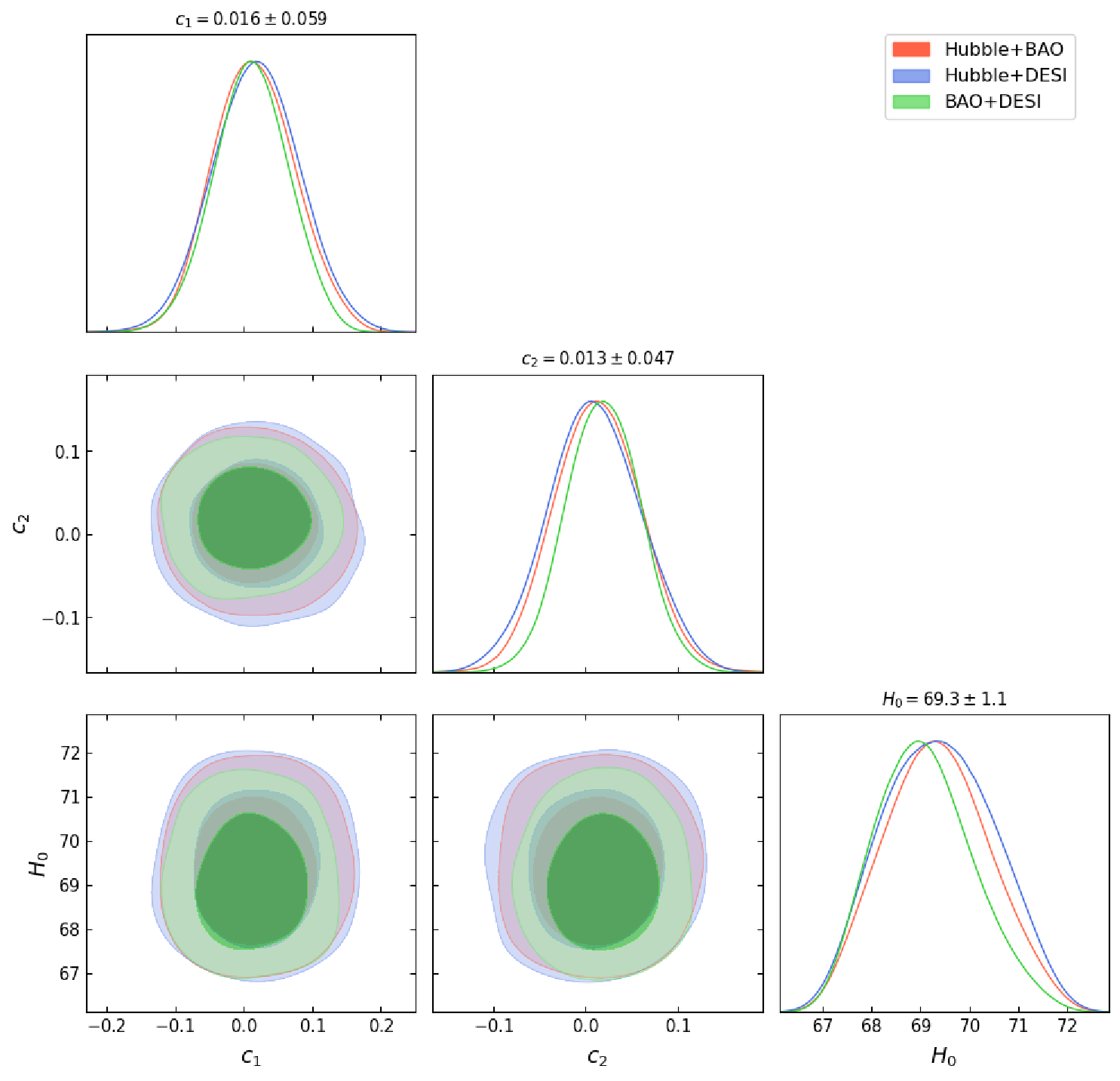}
    \caption{Joint and marginalized posterior distributions  with combinations of  datasets combination of  Hubble+BAO, Hubble+DESI, BAO+DESI, for Conventional Polynomial Parametrization}
    \label{fig:placeholder}
\end{figure}

\begin{figure}[h!]
    \centering
    \includegraphics[width=\linewidth]{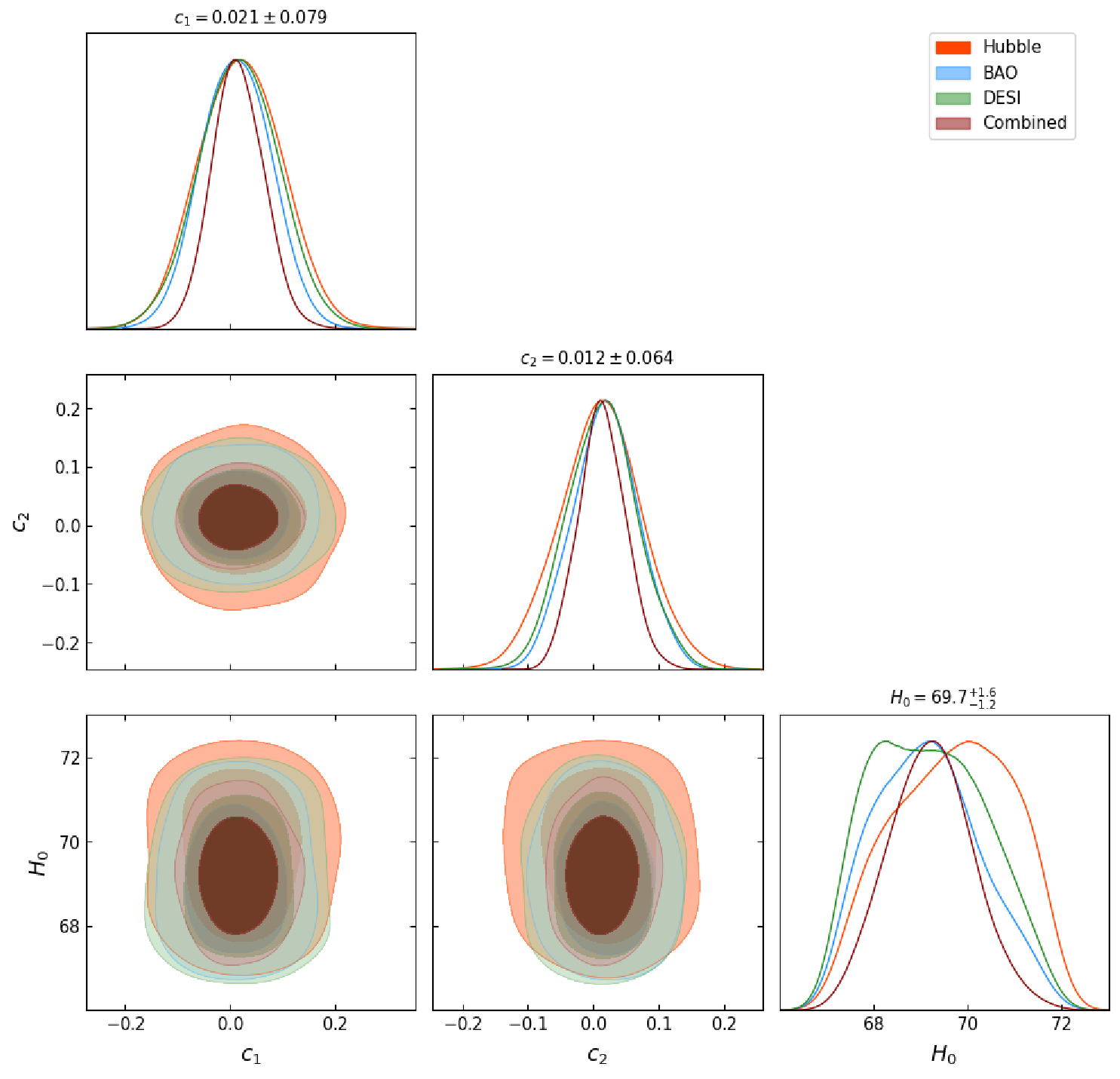}
    \caption{Joint and marginalized posterior distributions  with combinations of  datasets Hubble, BAO, DESI and Combined(Hubble+BAO+DESI) for Conventional Polynomial Parametrization}
    \label{fig:placeholder}
\end{figure}


\subsection{Legendre Polynomial Parametrization}
Using eqns.(\ref{hubb}) and (\ref{legendre}), we write the dimensionless Hubble parameter for the Legendre polynomial parametrization as
\begin{equation}
\frac{H^{2}(z)}{H_{0}^{2}}=\Omega_{m}(1+z)^{3}+\Omega_{de}(1+z)^{\frac{3}{8}(4c_{3}-c_{4})}(1+z^{2})^{\frac{3}{16}(4c_{3}+9c_{4})}
\exp\left[\frac{9c_{4}z(z-1)}{8(1+z^{2})}+\frac{3}{2}(c_{3}+3c_{4})\tan^{-1}(z)\right]
\end{equation}
This equation acts as the theoretical basis for this model. In Table 2, we have presented the best fit values of the free parameters $c_{3}$ and $c_{4}$ for the Legendre parametrization, along with the constrained values of the Hubble parameter $H_{0}$ for different combinations of datasets. In Fig.(4) we have presented the confidence contours showing the joint and marginalized posterior distribution with different datasets. In Fig.(5) similar confidence contours are presented for different combinations of the datasets taken two at a time. Finally, in Fig.(6) confidence contours are given for combined Hubble+BAO+DESI datasets.

\begin{table}[ht]
\centering
\caption{Parameter Constraints Summary for Legendre Polynomial Parametrization}
\begin{tabular}{lccc}
\hline
\hline
Dataset & $c_3$ & $c_4$ & $H_0$ [km/s/Mpc] \\
\hline
HUBBLE            & $0.0166 \pm 0.0809$ & $0.0080 \pm 0.0622$ & $69.65 \pm 1.23$ \\
BAO               & $0.0107 \pm 0.0674$ & $0.0208 \pm 0.0507$ & $69.13 \pm 1.18$ \\
DESI              & $0.0133 \pm 0.0748$ & $0.0148 \pm 0.0560$ & $69.23 \pm 1.22$ \\
HUBBLE+BAO        & $0.0151 \pm 0.0607$ & $0.0171 \pm 0.0462$ & $69.31 \pm 1.05$ \\
HUBBLE+DESI       & $0.0191 \pm 0.0662$ & $0.0109 \pm 0.0496$ & $69.45 \pm 1.11$ \\
BAO+DESI          & $0.0082 \pm 0.0561$ & $0.0198 \pm 0.0387$ & $69.07 \pm 0.99$ \\
HUBBLE+BAO+DESI   & $0.0143 \pm 0.0500$ & $0.0139 \pm 0.0356$ & $69.21 \pm 0.87$ \\
\hline
\multicolumn{4}{l}{\footnotesize Note: $\Omega_m = 0.31$ and $\Omega_{de} = 0.69$ are fixed parameters in this analysis.}
\end{tabular}
\label{tab:parameter_constraints}
\end{table}

\begin{figure}[h!]
    \centering
    \includegraphics[width=\linewidth]{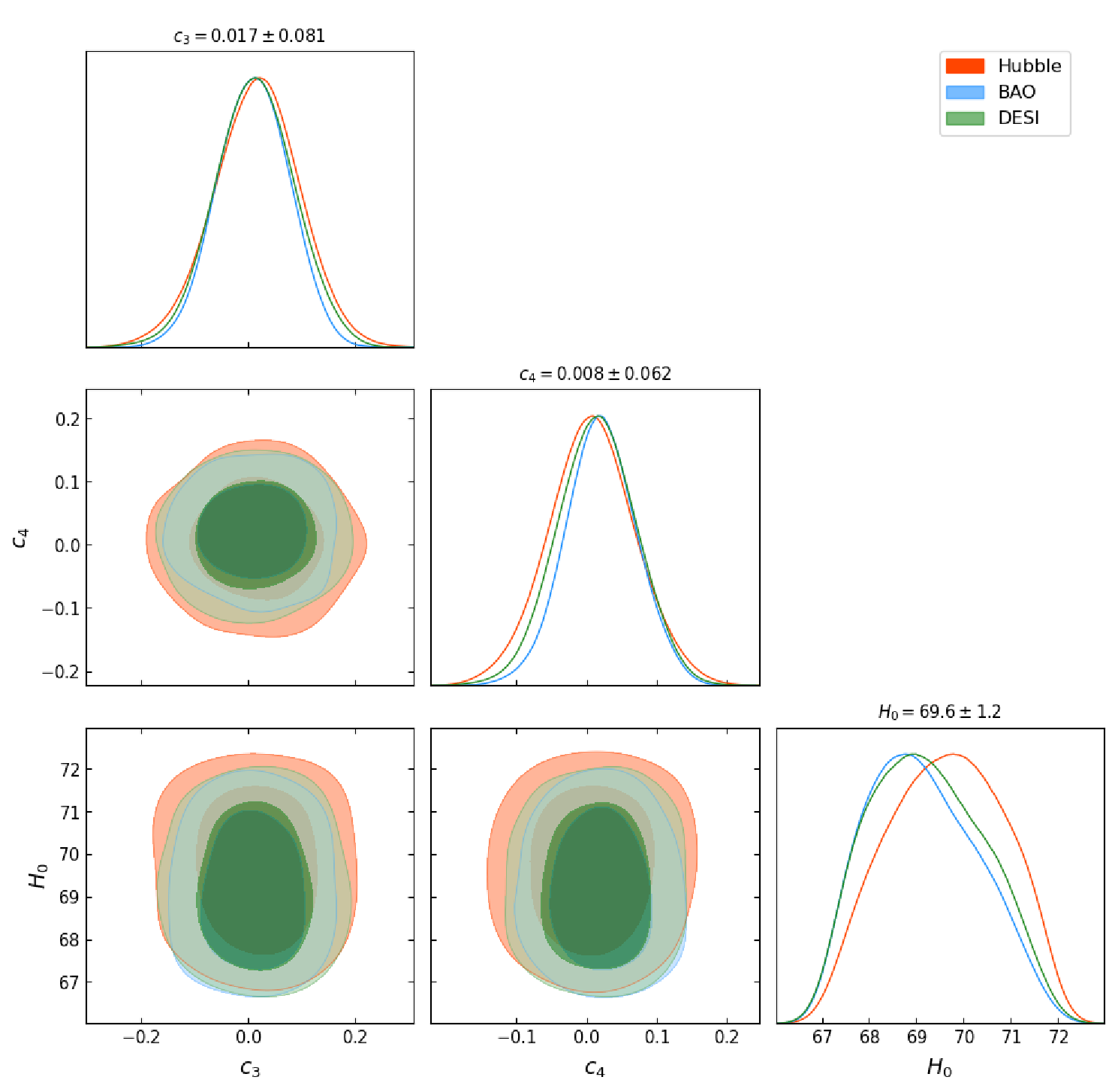}
    \caption{Joint and marginalized posterior distributions with combinations of datasets Hubble, BAO and DESI for Legendre Polynomial Parametrization}
    \label{fig:placeholder}
\end{figure}

\begin{figure}[h!]
    \centering
    \includegraphics[width=\linewidth]{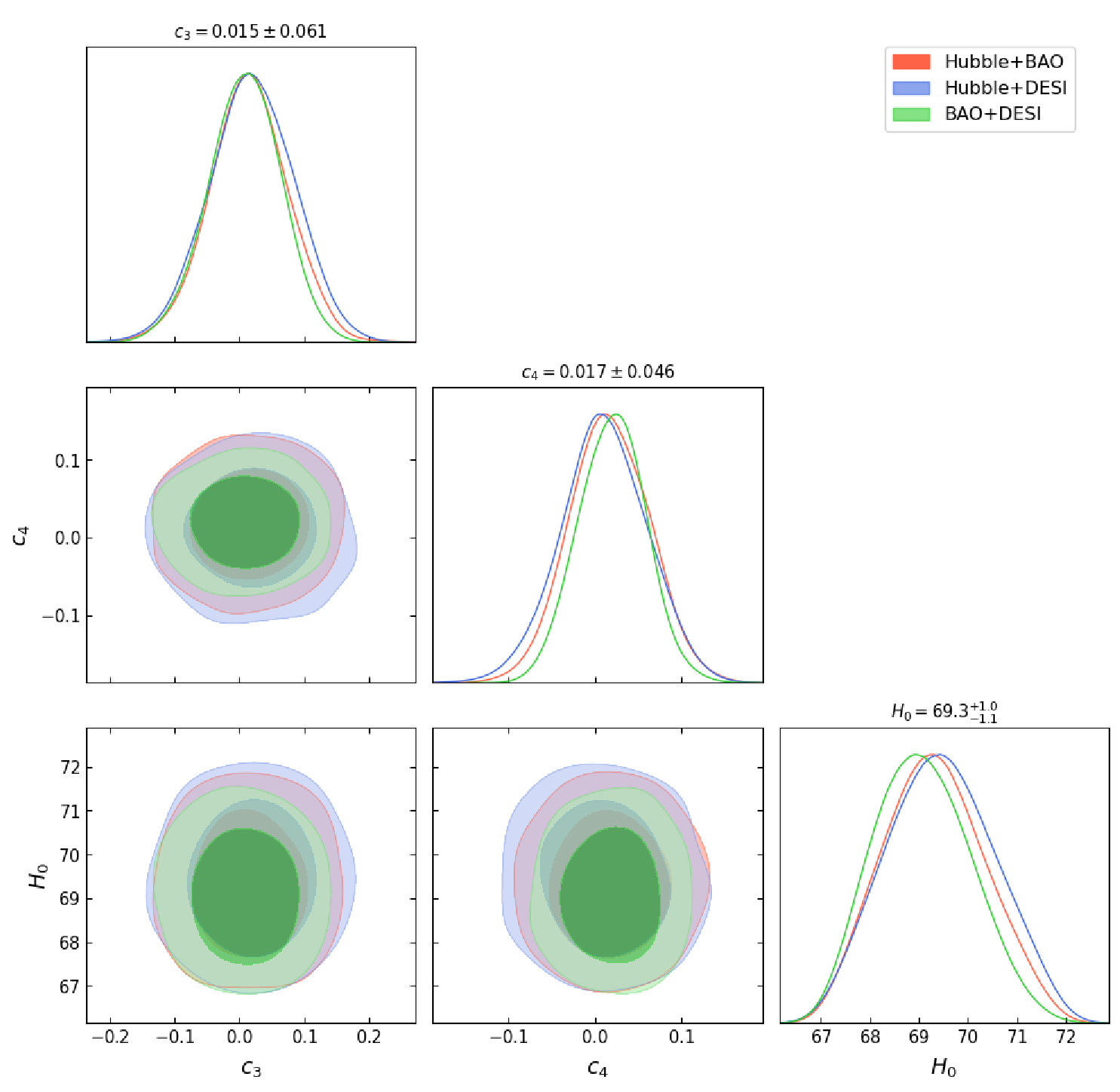}
    \caption{Joint and marginalized posterior distributions with combinations of datasets Hubble+BAO, Hubble+DESI, BAO+DESI for Legendre Polynomial Parametrization}
    \label{fig:placeholder}
\end{figure}

\begin{figure}[h!]
    \centering
    \includegraphics[width=\linewidth]{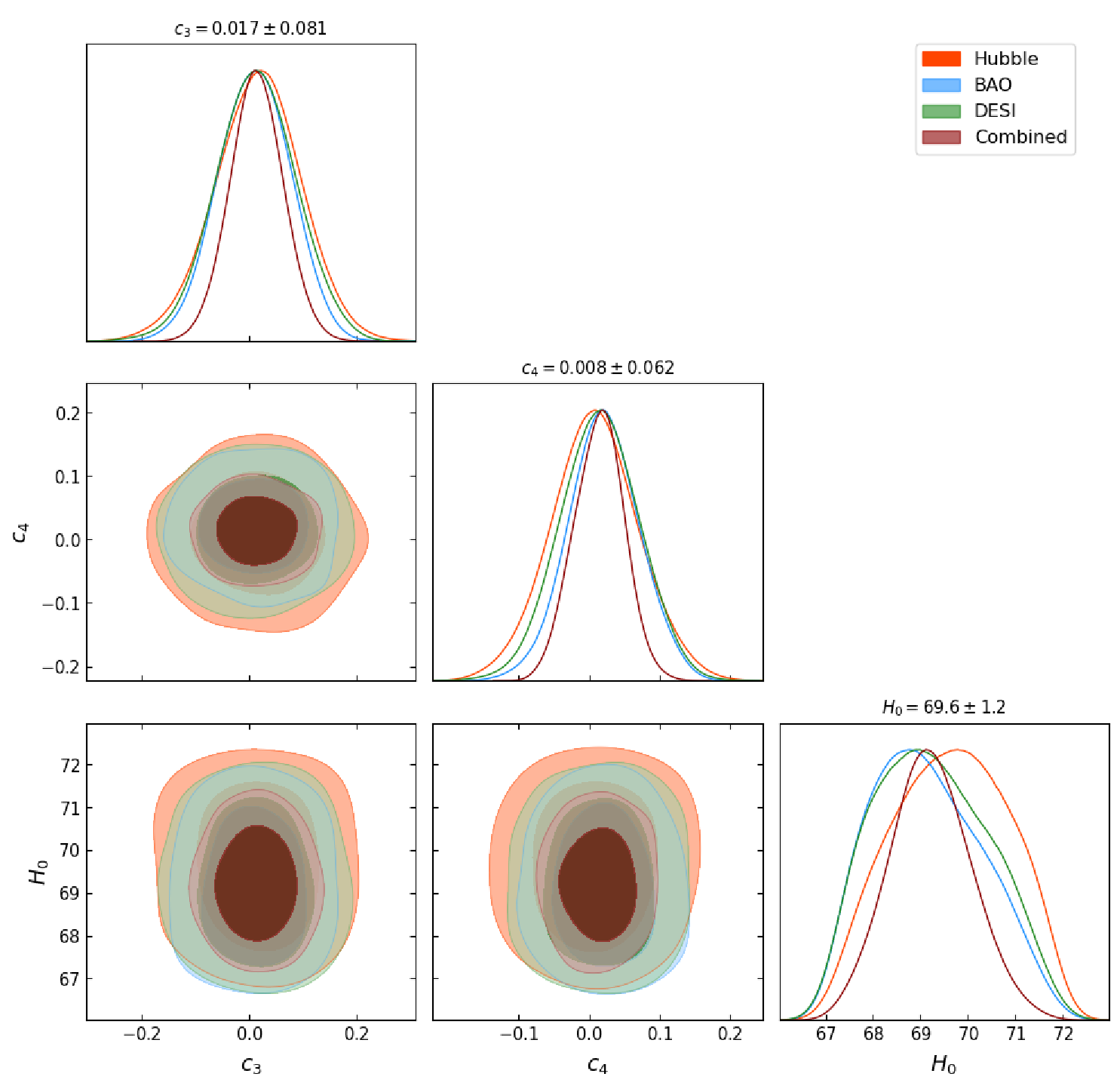}
    \caption{Joint and marginalized posterior distributions  with combinations of datasets Hubble, BAO, DESI and Combined(Hubble+BAO+DESI) for Legendre Polynomial Parametrization}
    \label{fig:placeholder}
\end{figure}

\subsection{Laguerre Polynomial Parametrization}
Using eqns.(\ref{hubb}) and (\ref{laguerre}), we write the dimensionless Hubble parameter for the Legendre polynomial parametrization as
\begin{equation}
\frac{H^{2}(z)}{H_{0}^{2}}=\Omega_{m}(1+z)^{3}+\Omega_{de}(1+z)^{\frac{3}{8}\left(4c_{5}+c_{6}\right)}(1+z^{2})^{-\frac{3}{16}\left(4c_{5}+5c_{6}\right)}
\exp\left[\frac{3}{8}\left(\frac{c_{6}z(z-1)}{1+z^{2}}-4\left(c_{5}+c_{6}\right)\tan^{-1}(z)\right)\right]
\end{equation}

In Table 3, we have presented the best fit values of the free parameters $c_{5}$ and $c_{6}$ for the Laguerre polynomial parametrization, along with the constrained values of the Hubble parameter $H_{0}$ for different combinations of datasets. In Fig. (7) we have presented the confidence contours showing the joint and marginalized posterior distribution with different datasets. In Fig.(8) similar confidence contours are presented for different combinations of the datasets taken two at a time. Finally, in Fig.(9) confidence contours are given for combined Hubble+BAO+DESI datasets.

\begin{table}[ht]
\centering
\caption{Parameter Constraints Summary for Laguerre Polynomial Parametrization}
\begin{tabular}{lccc}
\hline
\hline
Dataset & $c_5$ & $c_6$ & $H_0$ [km/s/Mpc] \\
\hline
HUBBLE            & $0.0177 \pm 0.0834$ & $0.0131 \pm 0.0594$ & $69.63 \pm 1.32$ \\
BAO               & $0.0125 \pm 0.0716$ & $0.0210 \pm 0.0490$ & $69.13 \pm 1.19$ \\
DESI              & $0.0147 \pm 0.0778$ & $0.0118 \pm 0.0545$ & $69.24 \pm 1.25$ \\
HUBBLE+BAO        & $0.0144 \pm 0.0629$ & $0.0160 \pm 0.0439$ & $69.38 \pm 1.12$ \\
HUBBLE+DESI       & $0.0220 \pm 0.0613$ & $0.0082 \pm 0.0473$ & $69.44 \pm 1.22$ \\
BAO+DESI          & $0.0112 \pm 0.0570$ & $0.0211 \pm 0.0409$ & $69.06 \pm 1.00$ \\
HUBBLE+BAO+DESI   & $0.0159 \pm 0.0495$ & $0.0161 \pm 0.0365$ & $69.19 \pm 0.90$ \\
\hline
\multicolumn{4}{l}{\footnotesize Note: $\Omega_m = 0.31$ and $\Omega_{de} = 0.69$ are fixed parameters in this analysis.}
\end{tabular}
\label{tab:parameter_constraints}
\end{table}

\begin{figure}[h!]
    \centering
    \includegraphics[width=\linewidth]{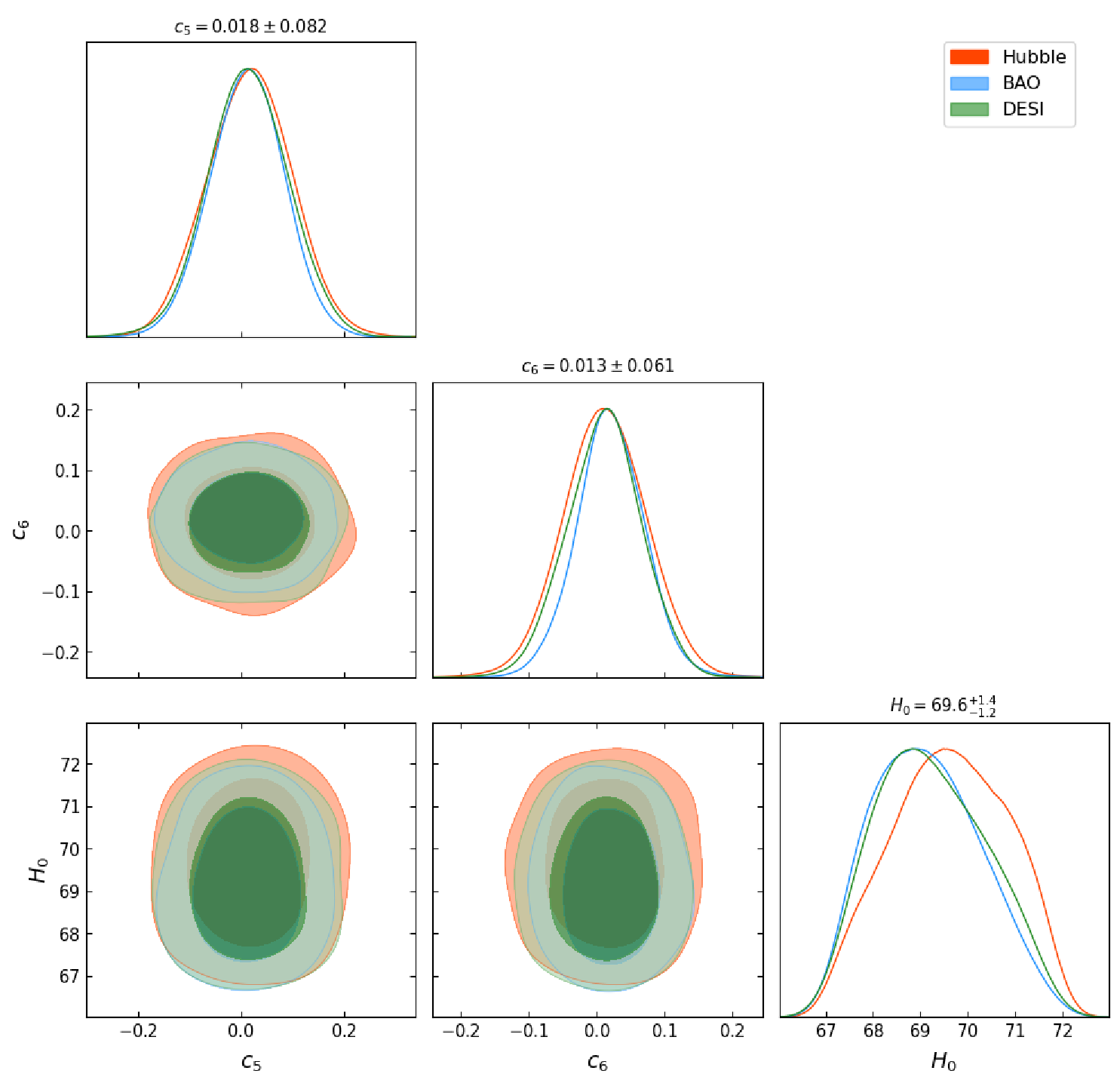}
    \caption{Joint and marginalized posterior distributions  with combinations of datasets Hubble, BAO and DESI for Laguerre Polynomial Parametrization}
    \label{fig:placeholder}
\end{figure}

\begin{figure}[h!]
    \centering
    \includegraphics[width=\linewidth]{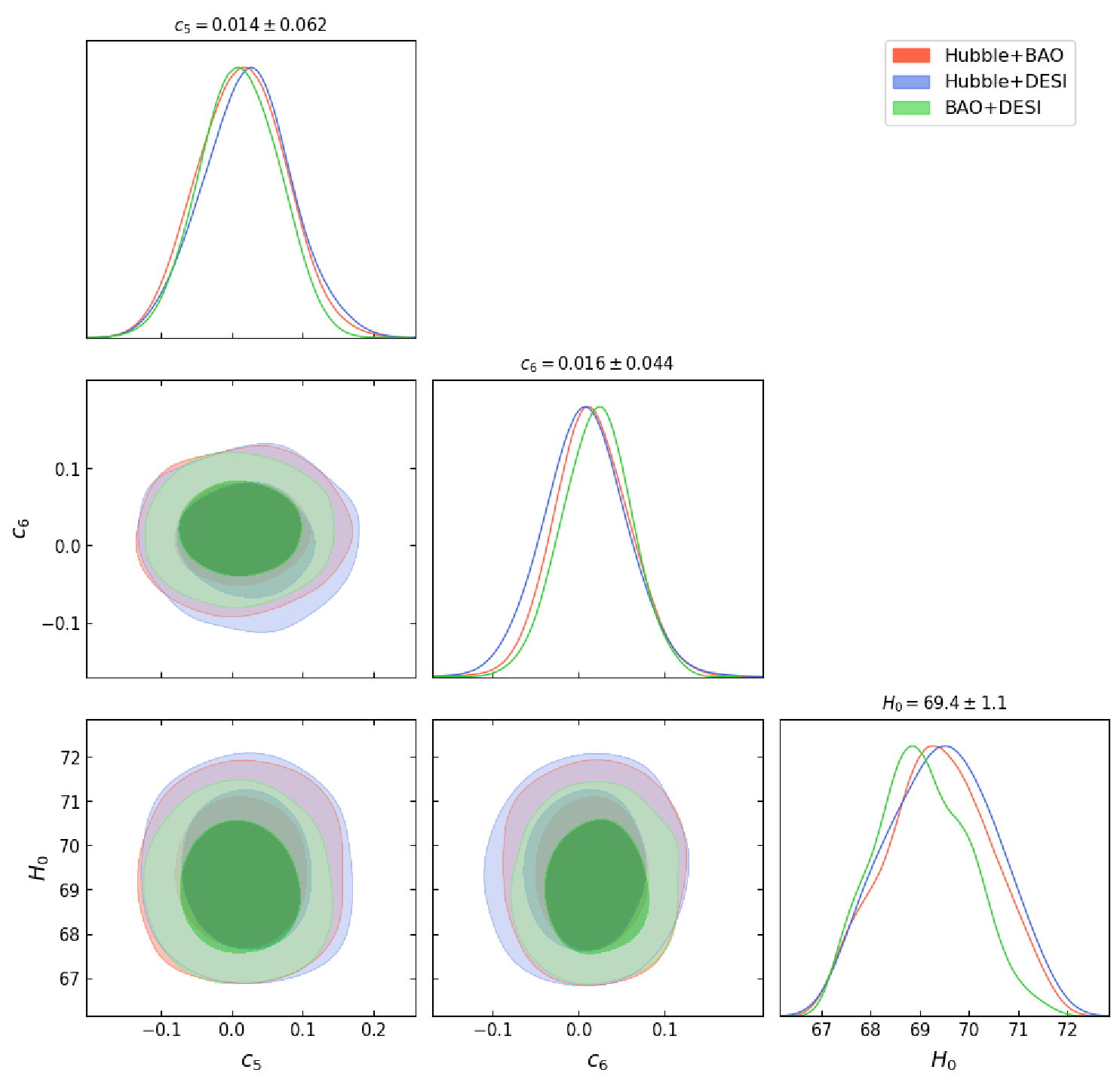}
    \caption{Joint and marginalized posterior distributions with combinations of datasets Hubble+BAO, Hubble+DESI, BAO+DESI, for Laguerre Polynomial Parametrization}
    \label{fig:placeholder}
\end{figure}

\begin{figure}[h!]
    \centering
    \includegraphics[width=\linewidth]{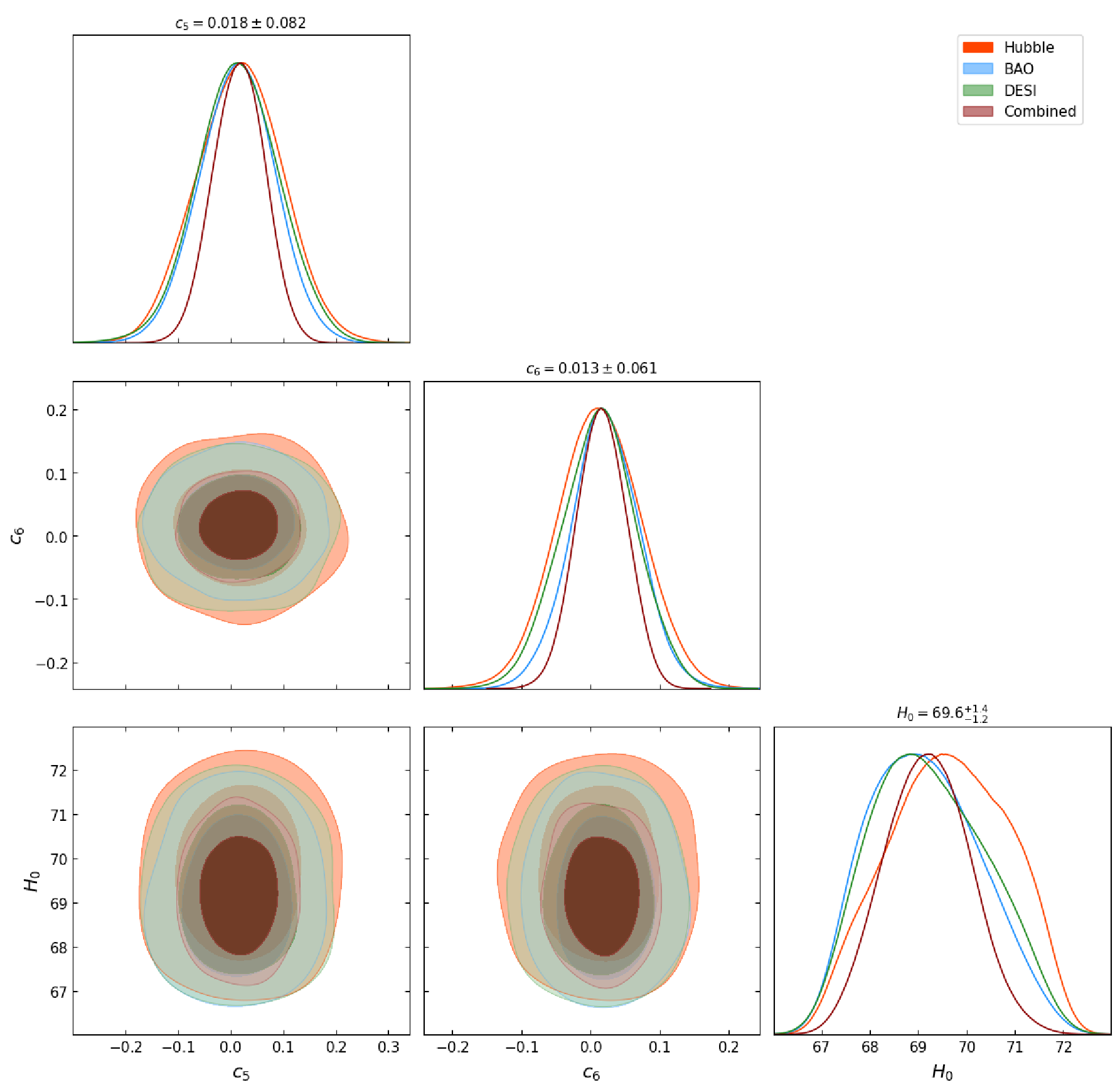}
    \caption{Joint and marginalized posterior distributions  with combinations of  datasets Hubble, BAO, DESI and Combined(Hubble+BAO+DESI) for Laguerre Polynomial Parametrization}
    \label{fig:placeholder}
\end{figure}

\subsection{Chebyshev Polynomial Parametrization}
Using eqns.(\ref{hubb}) and (\ref{chebyshev}), we write the dimensionless Hubble parameter for the Chebyshev polynomial parametrization as
\begin{equation}
\frac{H^{2}(z)}{H_{0}^{2}}=\Omega_{m}(1+z)^{3}+\Omega_{de}(1+z)^{\frac{3}{2}(c_{7}-c_{8})}(1+z^{2})^{\frac{3}{4}(c_{7}+3c_{8})}
\exp\left[\frac{3}{2}\left(\frac{c_{8}z(z-1)}{1+z^{2}}+(c_{7}+4c_{8})\tan^{-1}(z)\right)\right]
\end{equation}

In Table 4, we have presented the best fit values of the free parameters $c_{7}$ and $c_{8}$ for the Chebyshev polynomial parametrization, along with the constrained values of the Hubble parameter $H_{0}$ for different combinations of datasets. In Fig.(10) we have presented the confidence contours showing the joint and marginalized posterior distribution with different datasets. In Fig.(11) similar confidence contours are presented for different combinations of the datasets taken two at a time. Finally, in Fig.(12) confidence contours are given for the combined Hubble+BAO+DESI datasets.

\begin{table}[ht]
\centering
\caption{Parameter Constraints Summary  for Chebyshev Polynomial Parametrization}
\begin{tabular}{lccc}
\hline
\hline
Dataset & $c_7$ & $c_8$ & $H_0$ [km/s/Mpc] \\
\hline
HUBBLE            & $0.0202 \pm 0.0833$ & $0.0101 \pm 0.0606$ & $69.70 \pm 1.40$ \\
BAO               & $0.0136 \pm 0.0705$ & $0.0179 \pm 0.0477$ & $69.12 \pm 1.24$ \\
DESI              & $0.0105 \pm 0.0730$ & $0.0165 \pm 0.0567$ & $69.25 \pm 1.31$ \\
HUBBLE+BAO        & $0.0156 \pm 0.0608$ & $0.0135 \pm 0.0445$ & $69.32 \pm 1.07$ \\
HUBBLE+DESI       & $0.0205 \pm 0.0648$ & $0.0103 \pm 0.0518$ & $69.39 \pm 1.20$ \\
BAO+DESI          & $0.0098 \pm 0.0570$ & $0.0184 \pm 0.0411$ & $69.04 \pm 1.01$ \\
HUBBLE+BAO+DESI   & $0.0123 \pm 0.0492$ & $0.0162 \pm 0.0345$ & $69.22 \pm 0.86$ \\
\hline
\multicolumn{4}{l}{\footnotesize Note: $\Omega_m = 0.31$ and $\Omega_{de} = 0.69$ are fixed parameters in this analysis.}
\end{tabular}
\label{tab:parameter_constraints}
\end{table}
\begin{figure}[h!]
    \centering
    \includegraphics[width=\linewidth]{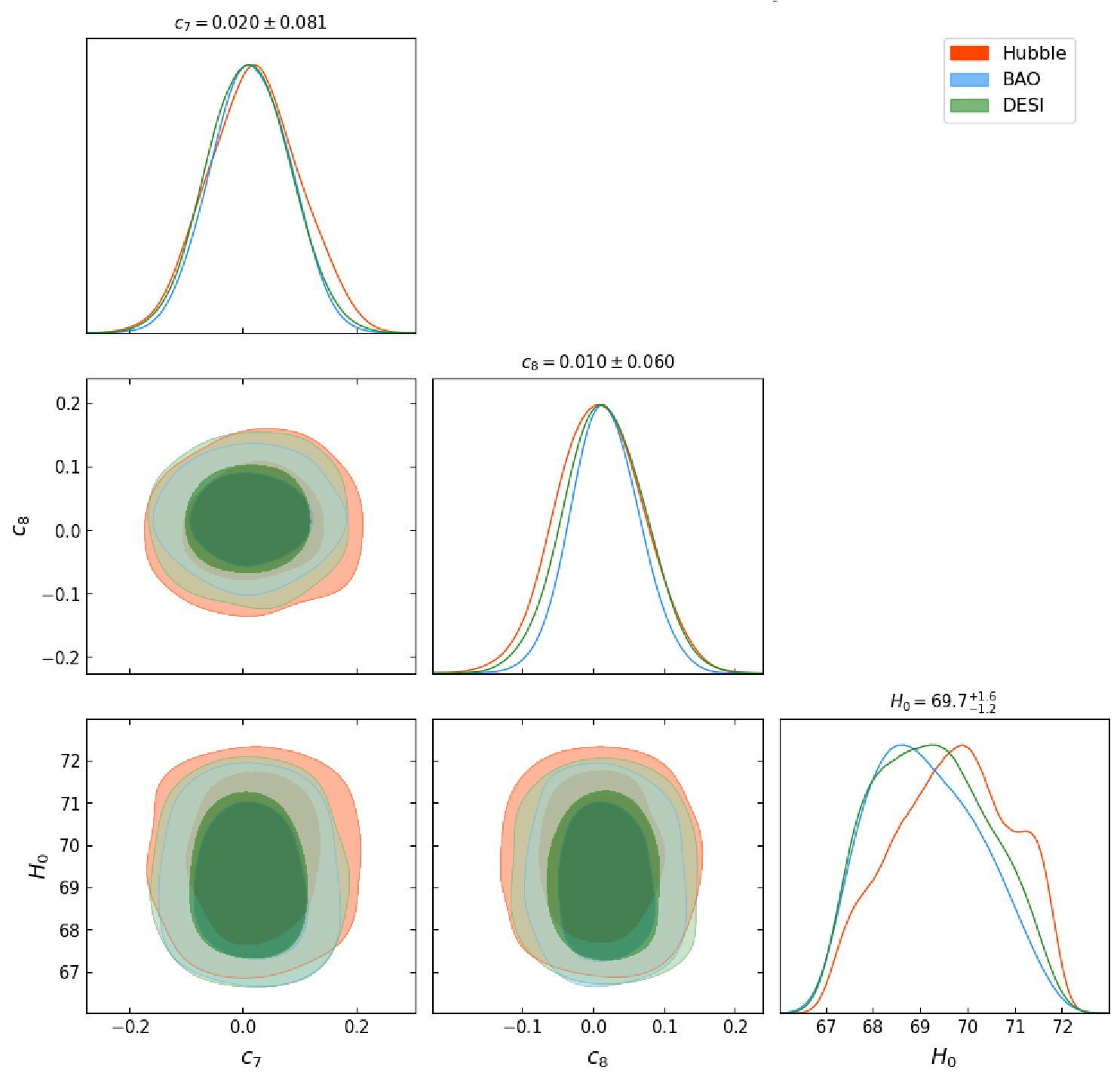}
    \caption{Joint and marginalized posterior distributions  with combinations of datasets Hubble, BAO and DESI for Chebyshev Polynomial Parametrization}
    \label{fig:placeholder}
\end{figure}

\begin{figure}[h!]
    \centering
    \includegraphics[width=\linewidth]{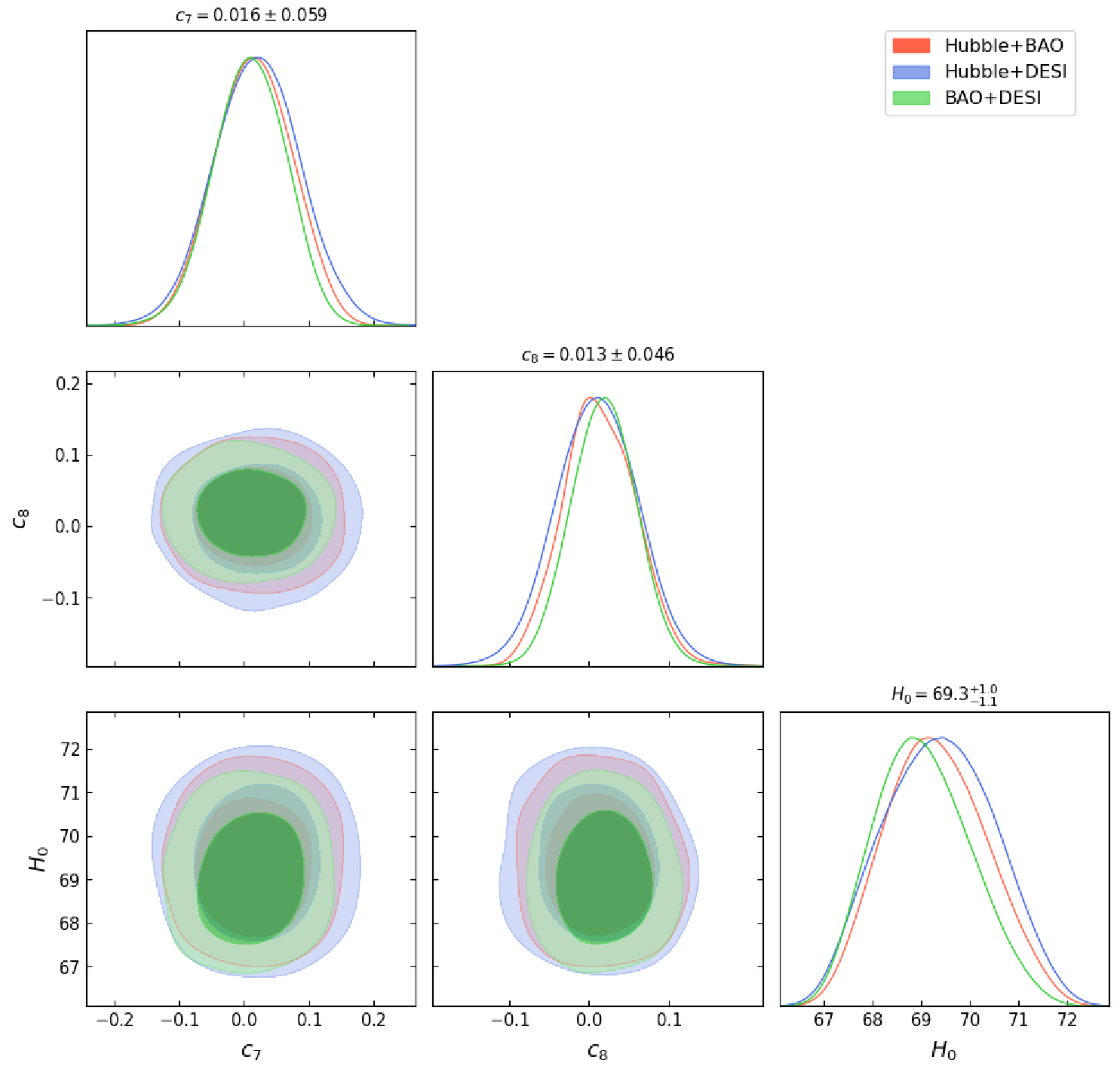}
    \caption{Joint and marginalized posterior distributions with combinations of datasets Hubble+BAO, Hubble+DESI, BAO+DESI for Chebyshev Polynomial Parametrization}
    \label{fig:placeholder}
\end{figure}

\begin{figure}[h!]
    \centering
    \includegraphics[width=\linewidth]{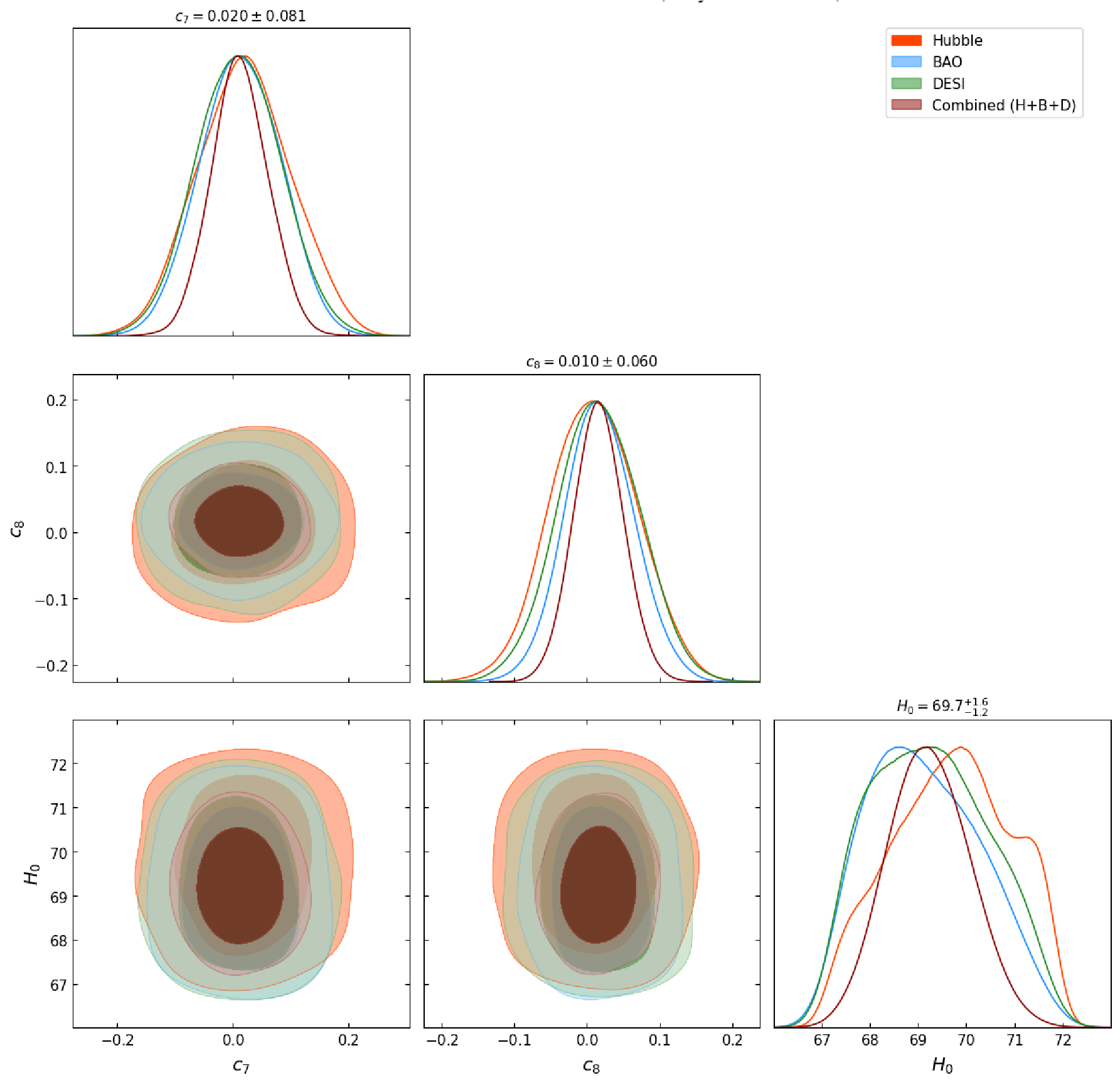}
    \caption{Joint and marginalized posterior distributions  with combinations of  datasets Hubble, BAO, DESI and Combined(Hubble+BAO+DESI) for Chebyshev Polynomial Parametrization}
    \label{fig:placeholder}
\end{figure}


\subsection{Fibonacci Polynomial Parametrization}
Using eqns.(\ref{hubb}) and (\ref{fibonacci}), we write the dimensionless Hubble parameter for the Fibonacci polynomial parametrization as
\begin{equation}
\frac{H^{2}(z)}{H_{0}^{2}}=\Omega_{m}(1+z)^{3}+\Omega_{de}
\exp\left[\frac{3}{4}\left(2c_{10}\tan^{-1}z+2(2c_{9}+c_{10})\ln(1+z)+c_{10}\ln(1+z^{2})\right)\right]
\end{equation}

In Table 5, we have presented the best fit values of the free parameters $c_{9}$ and $c_{10}$ for the Fibonacci polynomial parametrization, along with the constrained values of the Hubble parameter $H_{0}$ for different combinations of datasets. In Fig.(13) we have presented the confidence contours showing the joint and marginalized posterior distribution with different datasets. In Fig.(14) similar confidence contours are presented for different combinations of the datasets taken two at a time. Finally, in Fig.(15) confidence contours are given for the combined Hubble+BAO+DESI datasets.

\begin{table}[ht]
\centering
\caption{Parameter Constraints Summary  for Fibonacci Polynomial Parametrization}
\begin{tabular}{lccc}
\hline
\hline
Dataset & $c_9$ & $c_{10}$ & $H_0$ [km/s/Mpc] \\
\hline
HUBBLE            & $0.0193 \pm 0.0799$ & $0.0115 \pm 0.0605$ & $69.59 \pm 1.26$ \\
BAO               & $0.0127 \pm 0.0688$ & $0.0210 \pm 0.0525$ & $69.17 \pm 1.18$ \\
DESI              & $0.0150 \pm 0.0763$ & $0.0156 \pm 0.0559$ & $69.18 \pm 1.20$ \\
HUBBLE+BAO        & $0.0142 \pm 0.0603$ & $0.0158 \pm 0.0456$ & $69.36 \pm 1.07$ \\
HUBBLE+DESI       & $0.0200 \pm 0.0647$ & $0.0088 \pm 0.0506$ & $69.45 \pm 1.13$ \\
BAO+DESI          & $0.0097 \pm 0.0557$ & $0.0200 \pm 0.0405$ & $69.03 \pm 0.99$ \\
HUBBLE+BAO+DESI   & $0.0135 \pm 0.0494$ & $0.0148 \pm 0.0335$ & $69.15 \pm 0.86$ \\
\hline
\multicolumn{4}{l}{\footnotesize Note: $\Omega_m = 0.31$ and $\Omega_{de} = 0.69$ are fixed parameters in this analysis.}
\end{tabular}
\label{tab:parameter_constraints}
\end{table}

\begin{figure}[h!]
    \centering
    \includegraphics[width=\linewidth]{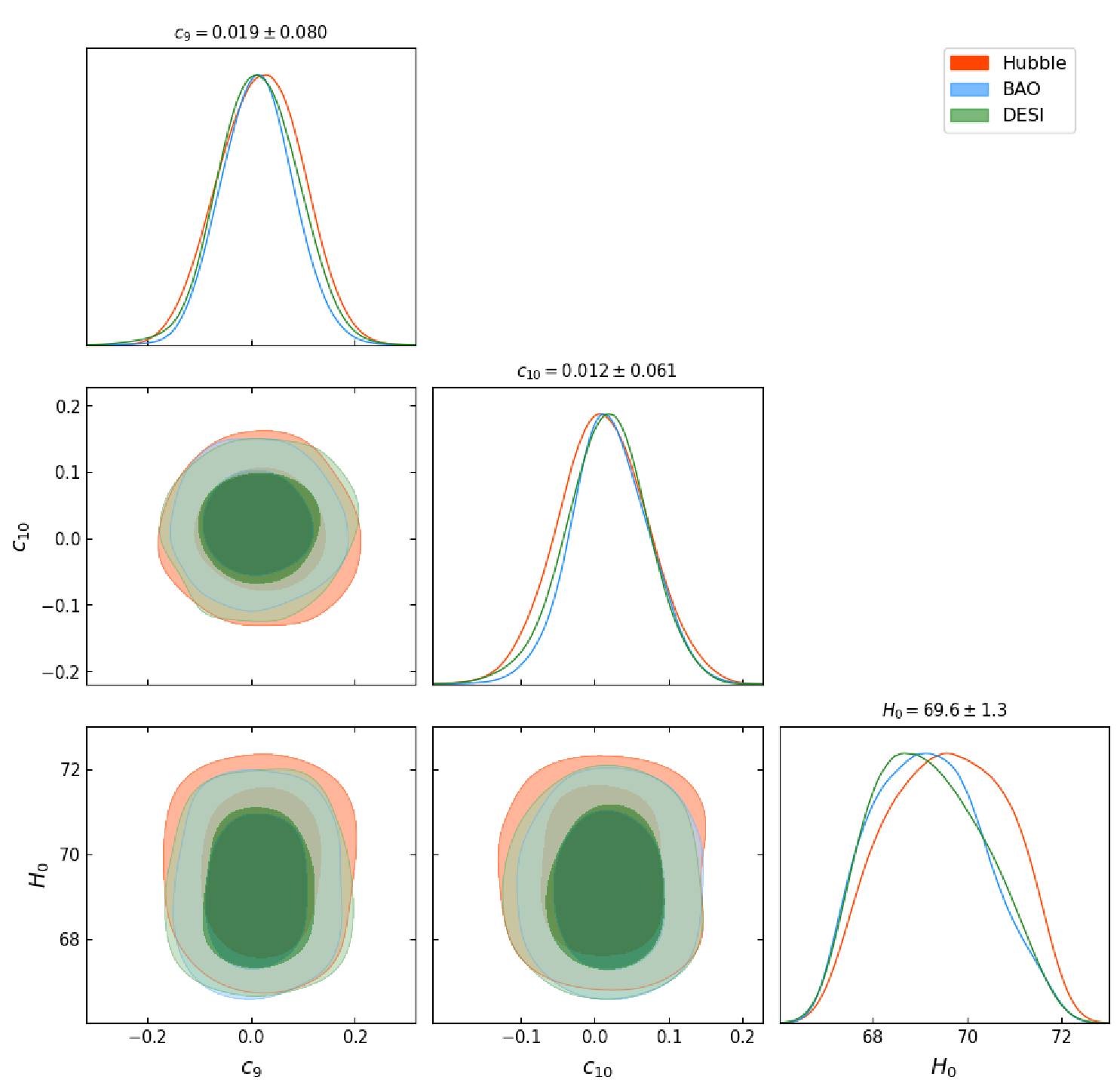}
    \caption{Joint and marginalized posterior distributions  with combinations of datasets Hubble, BAO and DESI for Fibonacci Polynomial Parametrization}
    \label{fig:placeholder}
\end{figure}

\begin{figure}[h!]
    \centering
    \includegraphics[width=\linewidth]{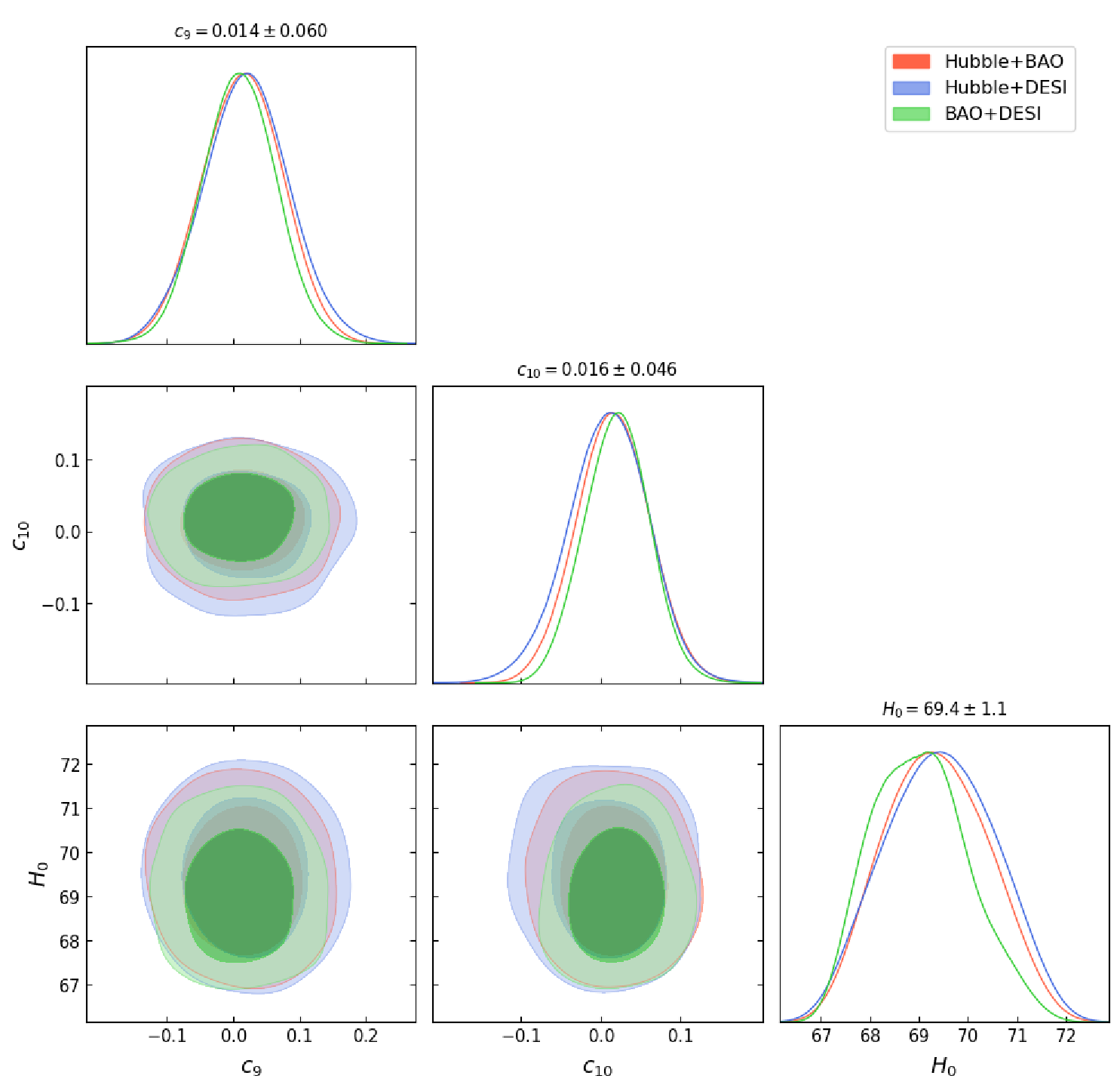}
    \caption{Joint and marginalized posterior distributions  with combinations of datasets Hubble+BAO, Hubble+DESI, BAO+DESI for Fibonacci Polynomial Parametrization}
    \label{fig:placeholder}
\end{figure}

\begin{figure}[h!]
    \centering
    \includegraphics[width=\linewidth]{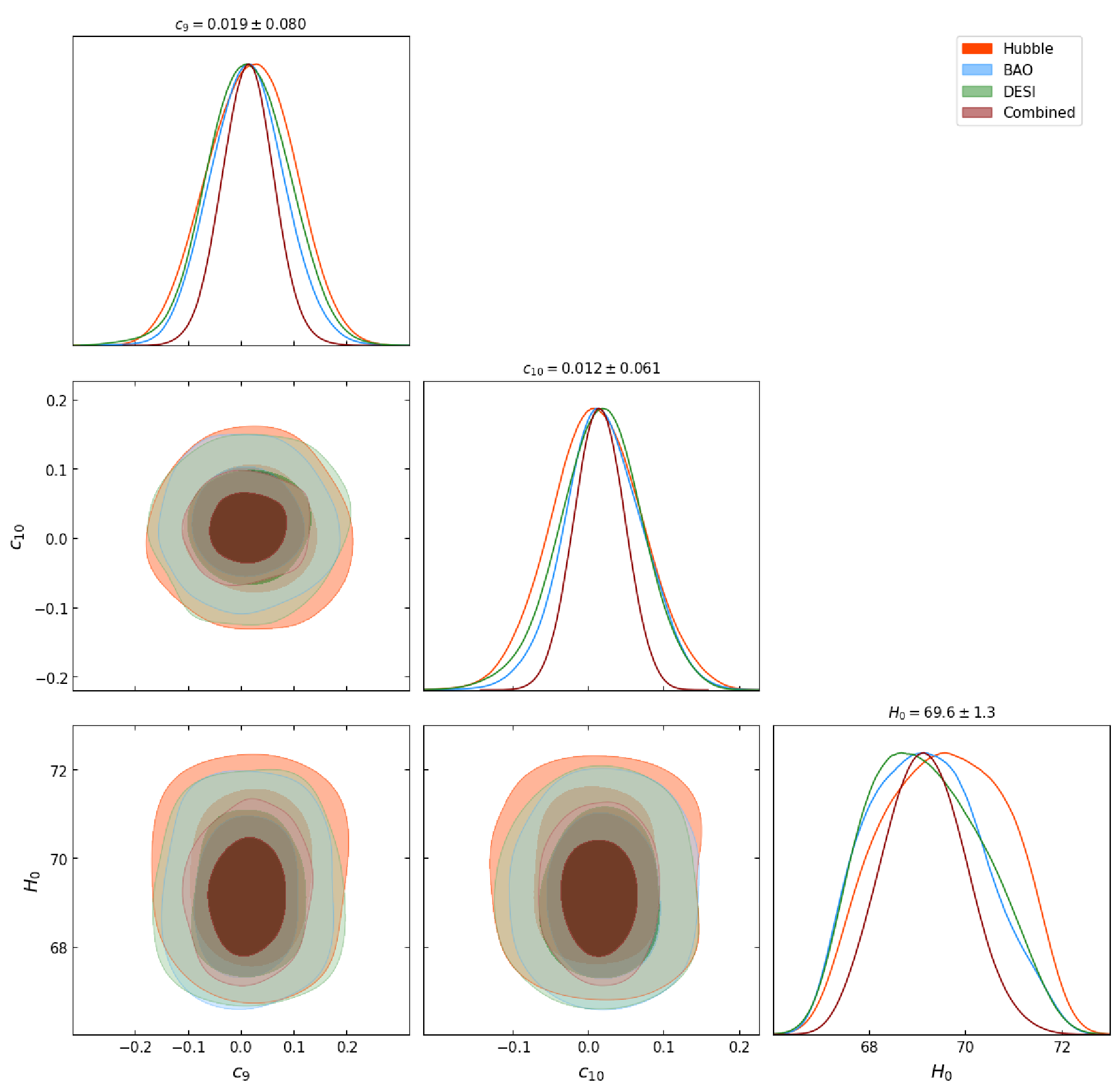}
    \caption{Joint and marginalized posterior distributions  with combinations of datasets Hubble, BAO, DESI and Combined(Hubble+BAO+DESI) for Fibonacci Polynomial Parametrization}
    \label{fig:placeholder}
\end{figure}

\section{Application of Machine Learning for model comparison}
We employed six supervised machine learning (ML) regression algorithms to reconstruct the Hubble parameter function and compare their performance with our theoretical model. The algorithms used include Enhanced Linear Regression (ELR), Physics-Informed Linear Regression (PILR), Enhanced Artificial Neural Network (ANN), Enhanced Support Vector Regression (SVR), Enhanced Random Forest Regression (ERFR), and Gradient Boosting Regression (GBR). These models were trained using a 20-point training set and evaluated on a 10-point test set. Their performance was assessed through the coefficient of determination ($R^2$), root mean square error (RMSE), chi-squared statistics, reduced chi-squared, and mean absolute deviation in the parameter $\alpha$. The comparative analysis of various ML and theoretical models under different polynomial bases—namely, Conventional, Legendre, Laguerre, Chebyshev, and Fibonacci—provides valuable insight into the numerical stability and physical consistency of the proposed modelling framework. Tables 6, 7, 8, 9, 10 and Figs.(16), (17), (18), (19), (20) present the results and intricacies of all the machine learning algorithms for the different redshift parametrizations.

Across all polynomial bases, the Enhanced Support Vector Regression (Enhanced\_SVR) exhibits consistently high predictive performance, maintaining an average training $R^2 \approx 0.93$ and test $R^2 \approx 0.87$. This stability demonstrates the SVR model’s strong capability to capture nonlinear relationships between the underlying physical parameters while preserving generalization to unseen data. Its moderate RMSE values (around 15.8) and low reduced $\chi^2$ ($\sim 0.49$) indicate a balanced trade-off between model complexity and sensitivity to noise. The Enhanced Artificial Neural Network (Enhanced\_ANN) performs comparably, achieving test $R^2$ values around 0.84–0.85. The close agreement between SVR and ANN results suggests that both nonlinear regression techniques effectively learn the physical mapping between inputs and target quantities. However, the slightly higher RMSE of the ANN (approximately 17.1) implies greater sensitivity to data dispersion, likely due to weight regularization and dependence on initialization parameters. The Physics-Informed Linear Model shows a distinct advantage in interpretability, embedding theoretical constraints directly into the learning process. Although its predictive accuracy is marginally lower than that of the Enhanced\_SVR, its reduced $\chi^2$ values (typically around 0.55) remain physically consistent, implying that the model’s residuals are within acceptable theoretical limits. The close alignment of its results under different basis transformations confirms that integrating physical constraints improves numerical stability and reduces overfitting to purely empirical patterns. In contrast, the Enhanced Linear Model displays slightly weaker generalization ($R^2_{\text{test}} \approx 0.82$) and higher RMSE ($\approx 18.17$), emphasizing that while linear structures capture broad trends, they struggle to reproduce subtle nonlinear dependencies inherent in cosmic dynamics. The elevated reduced $\chi^2$ values ($\sim 0.63$) and larger mean $\alpha$ deviations further support this limitation. The ensemble-based algorithms, namely Enhanced Random Forest and Gradient Boosting, achieve very high training accuracy ($R^2_{\text{train}} > 0.96$) but show a notable drop in test performance ($R^2_{\text{test}} \sim 0.78$), indicating mild overfitting. This behavior is characteristic of tree-based methods, where deep branching can model noise rather than the underlying physical signal. Despite this, both ensemble learners yield reasonably good reduced $\chi^2$ values (around 0.60–0.68), suggesting that ensemble averaging helps mitigate overfitting. Their moderate mean $\alpha$ deviations (below 0.08) confirm good local reliability but lower robustness when extrapolating beyond the training range. A particularly noteworthy outcome arises from the Modified Differential Evolution (MDE) Theoretical Model, which consistently outperforms all empirical and hybrid approaches in both error minimization and physical consistency. The MDE-based model achieves the lowest RMSE and reduced $\chi^2$ across all basis expansions. For example, under the Legendre and Chebyshev bases, it yields $\text{RMSE}_{\text{Legendre}} = 11.8788$ with $\text{Reduced } \chi^2_{\text{Legendre}} = 0.4699$, and $\text{RMSE}_{\text{Chebyshev}} = 11.8810$ with $\text{Reduced } \chi^2_{\text{Chebyshev}} = 0.4701$. These nearly identical values indicate high numerical convergence and physical robustness. The MDE approach, which optimizes a theoretically motivated objective function rather than relying solely on data-driven fitting, demonstrates superior adaptability in minimizing residual variance across different orthogonal representations. Its physically constrained optimization ensures that model parameters respect boundary conditions, leading to a more accurate reconstruction of the underlying physical law. A comparison across different basis functions reveals a clear pattern in how orthogonality influences physical accuracy. The Legendre and Chebyshev bases, both orthogonal polynomial systems, yield the most stable and precise results. Their mathematical conditioning minimizes multicollinearity among terms, thereby reducing numerical error propagation. Consequently, these bases produce the lowest RMSE and $\chi^2$ values, confirming their suitability for representing smooth, physically continuous relations. The Laguerre and Fibonacci bases also perform effectively but exhibit slightly higher RMSE values (around 12.0), suggesting that non-orthogonal or quasi-orthogonal expansions introduce minor instabilities in higher-order terms. The Conventional basis shows relatively weaker performance, highlighting the advantage of transitioning from simple polynomial to orthogonal polynomial frameworks in reconciling data with theory. In conclusion, integrating physically informed structures with orthogonal basis expansions significantly enhances both predictive accuracy and theoretical fidelity. The MDE theoretical Model, particularly in combination with Legendre or Chebyshev bases, emerges as the most statistically robust and physically consistent formulation. The low reduced $\chi^2$ ($\approx 0.47$) and minimal mean $\alpha$ deviation ($\sim 0.078$) confirm that this hybrid theoretical–numerical approach captures the essential dynamics of the system with high precision. Overall, this comprehensive analysis demonstrates that employing orthogonal polynomial representations within physics-informed or DE-optimized schemes greatly strengthens the reliability, interpretability, and extrapolative capacity of modern physical modelling frameworks.

\begin{figure}[h!]
    \centering
    \includegraphics[width=0.48\linewidth]{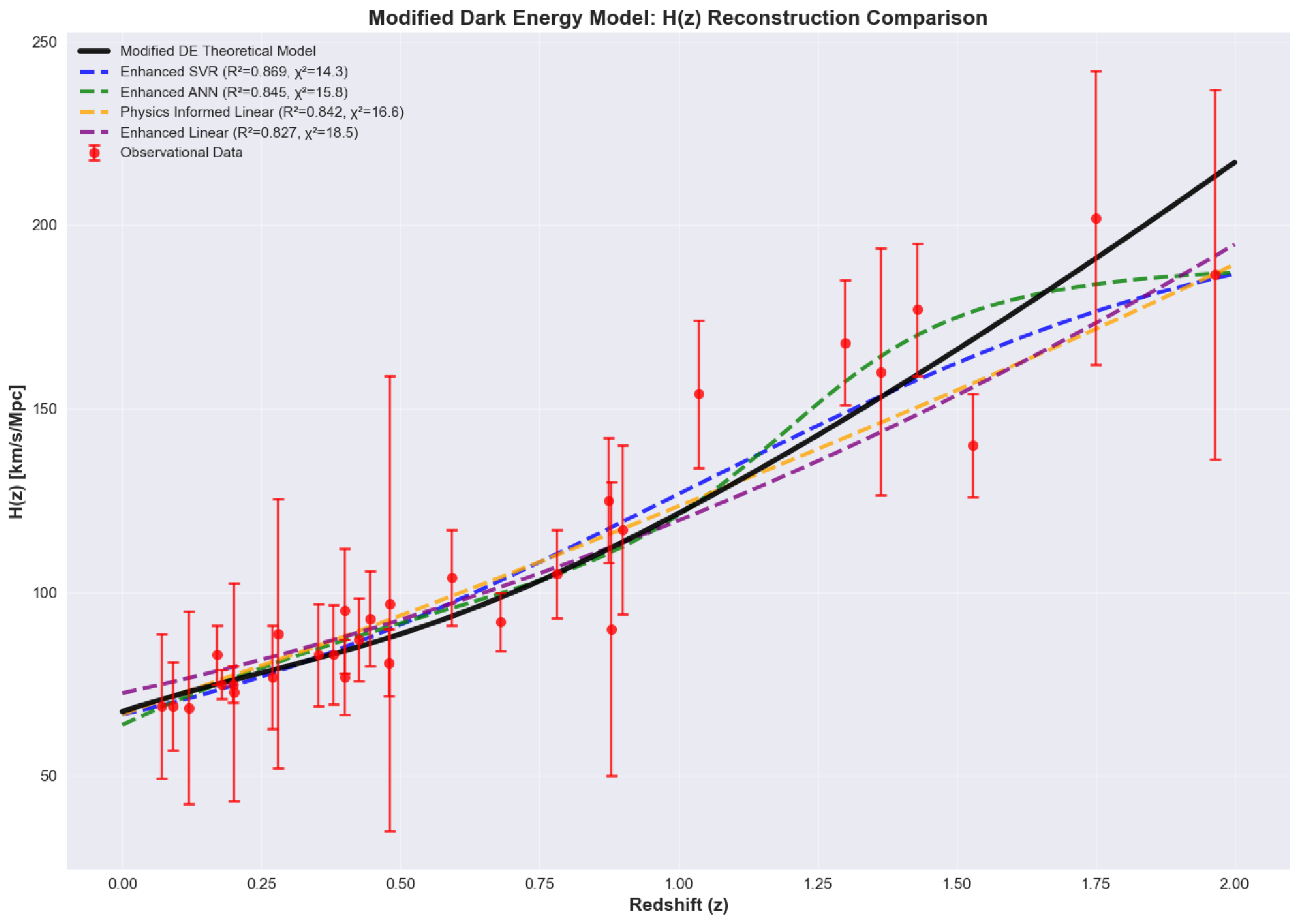}
    \hfill
    \includegraphics[width=0.48\linewidth]{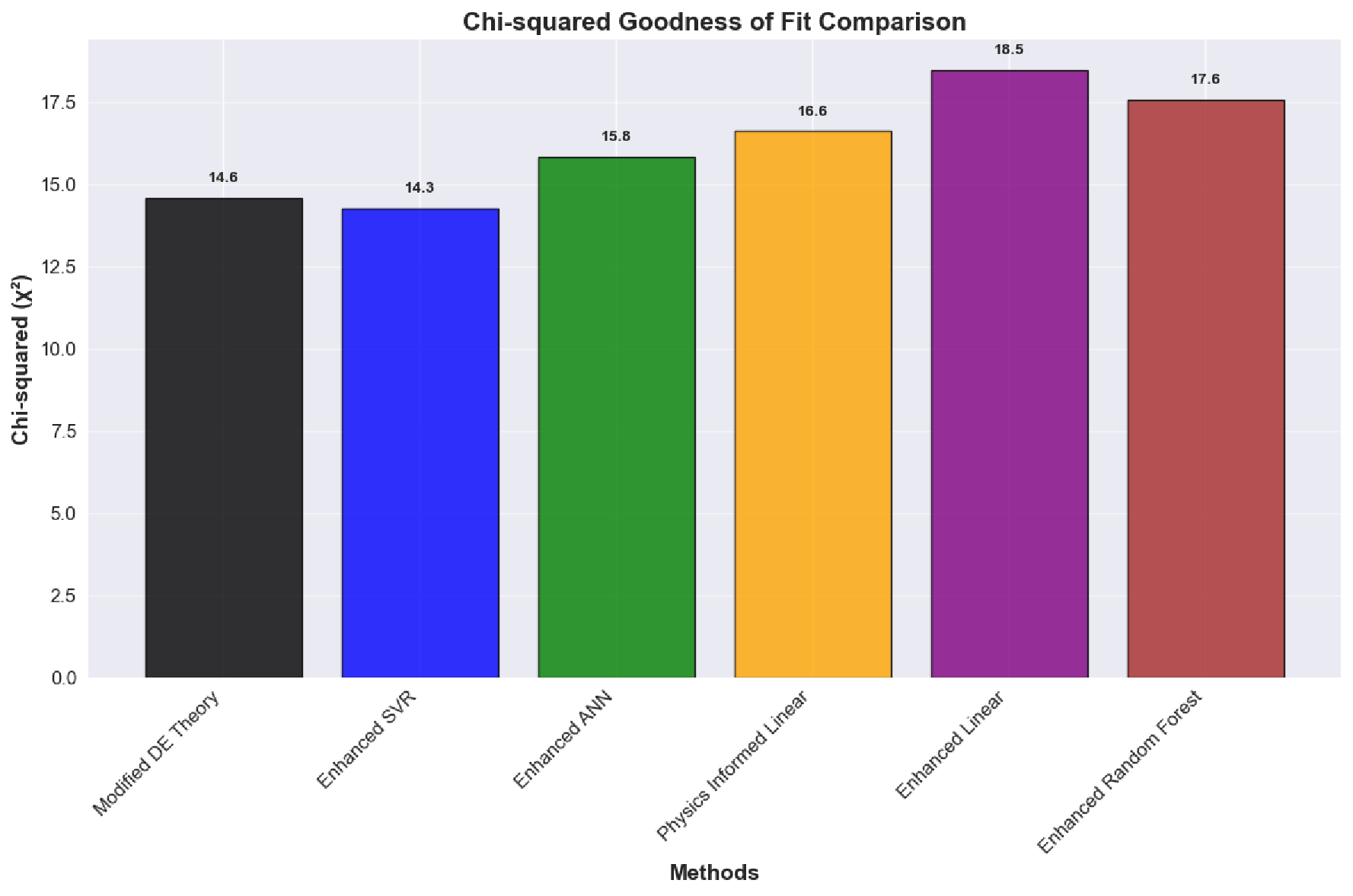}
    \\[0.5cm]
    \includegraphics[width=0.58\linewidth]{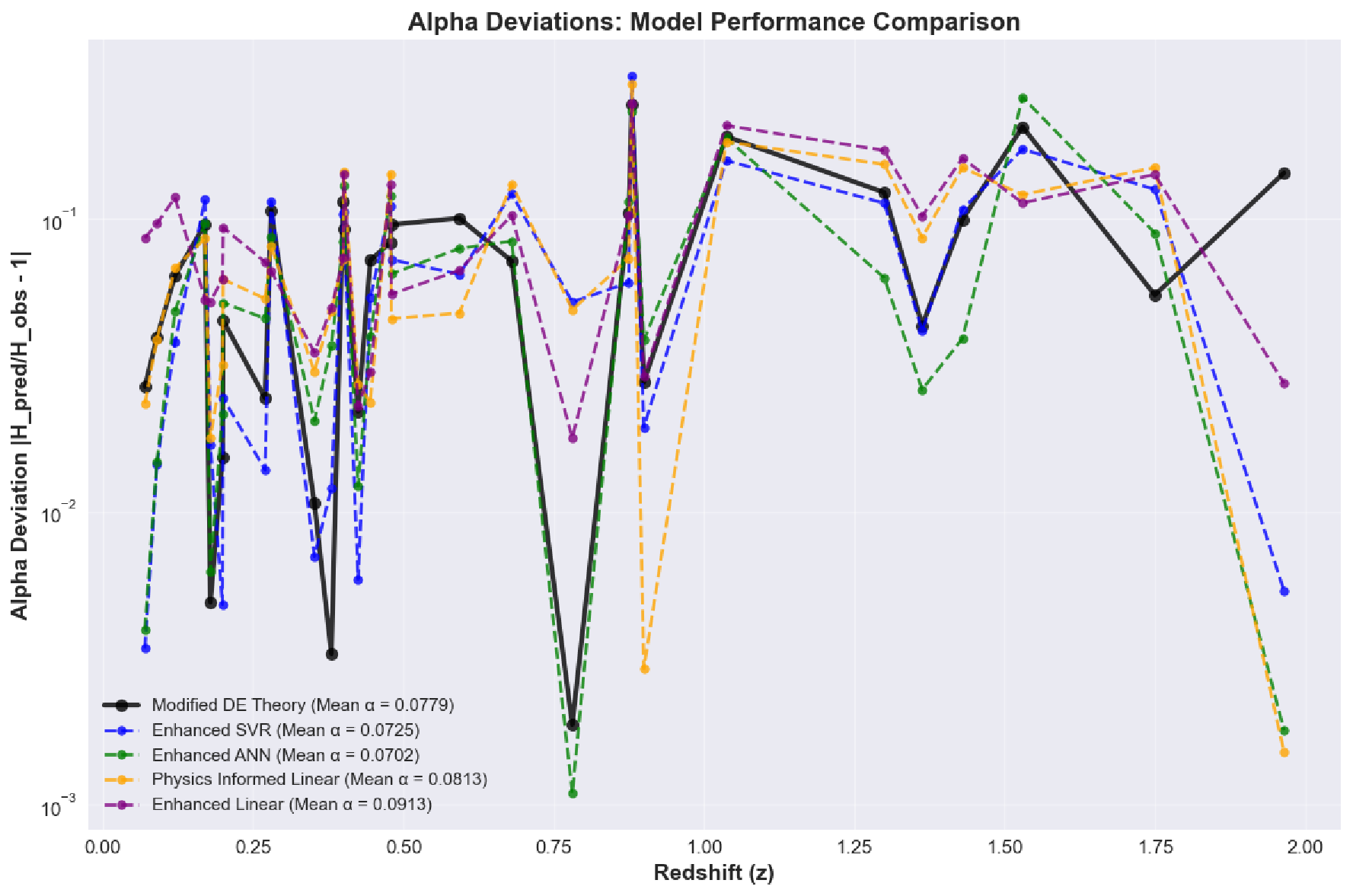}
    \caption{(Left) Machine learning regression fit to observational Hubble parameter $H(z)$ data using our model. The solid black line shows the theoretical prediction, while data points (red circles) correspond to measured $H(z)$ values with $1\sigma$ error bars, the regression fits from ML algorithms are shown using dashed lines, and the close alignment illustrates the model's accuracy in capturing the cosmic expansion history for the Conventional model. (Right) $\chi^2$ goodness of fit for different ML techniques in a bar diagram showing their comparative values for the conventional model. (Down) Comparison of the absolute relative deviation $|H_{\mathrm{pred}}/H_{\mathrm{obs}} - 1|$ of the Hubble parameter from different predictive models as a function of redshift. The curves represent the theory, Enhanced Support Vector Regression (SVR), Enhanced Artificial Neural Network (ANN), Enhanced Linear Regression, and Physics-Informed Linear Regression. The plot demonstrates that machine learning models, especially Enhanced SVR and ANN, exhibit deviations comparable to the theoretical model, indicating strong predictive performance for the conventional model.}
    \label{fig:conventional}
\end{figure}

\begin{figure}[h!]
    \centering
    \includegraphics[width=0.48\linewidth]{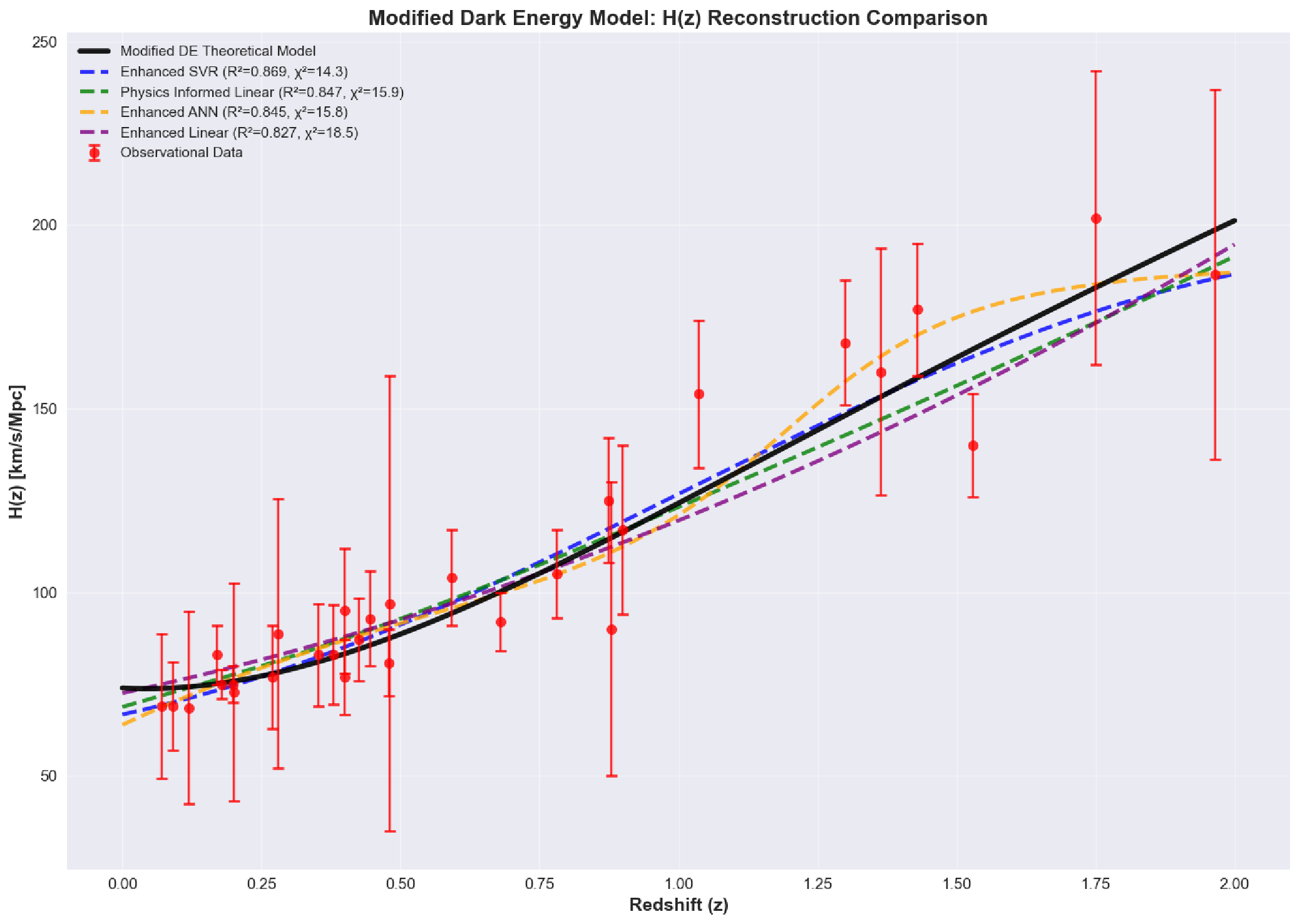}
    \hfill
    \includegraphics[width=0.48\linewidth]{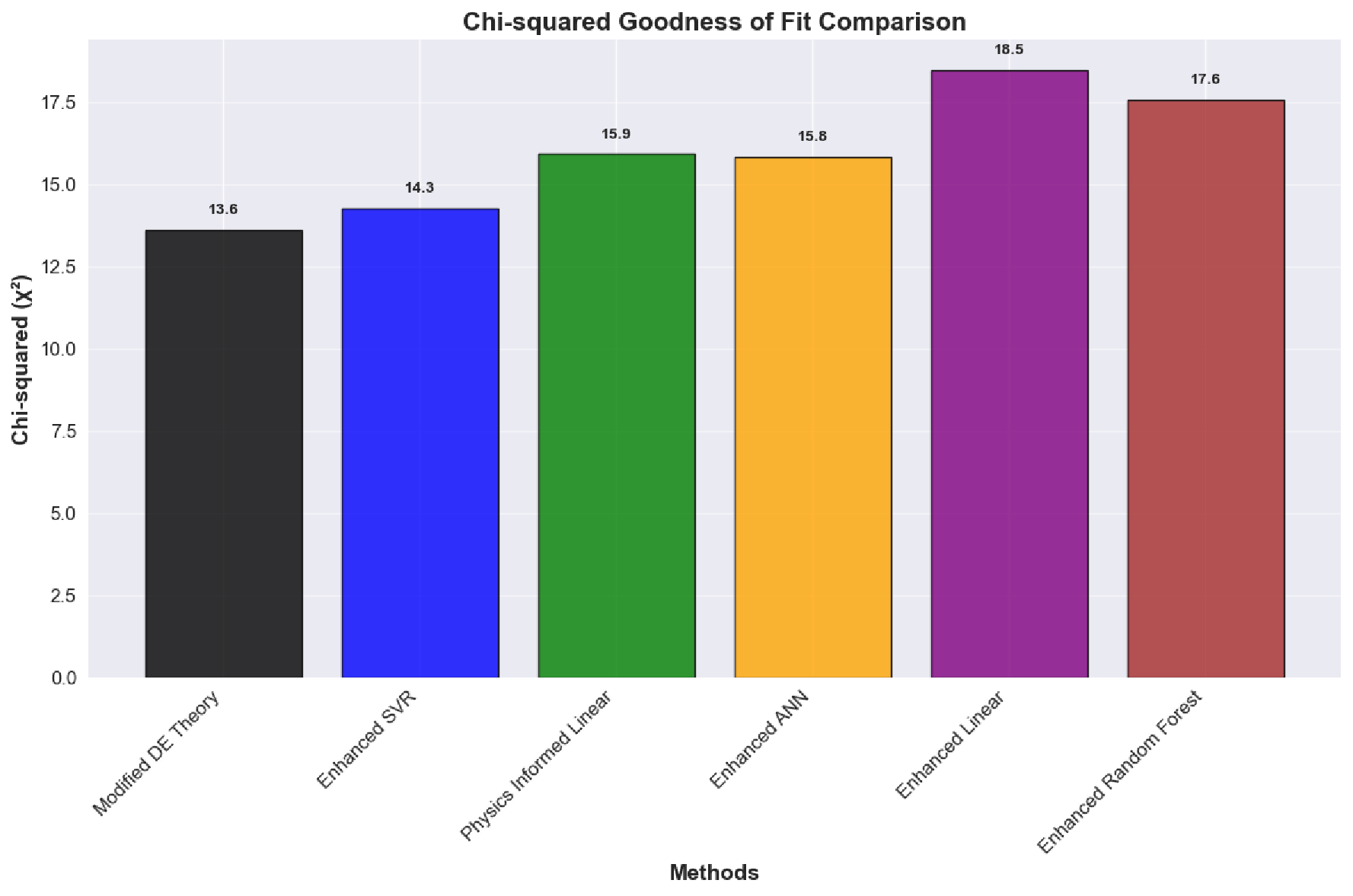}
    \\[0.5cm]
    \includegraphics[width=0.58\linewidth]{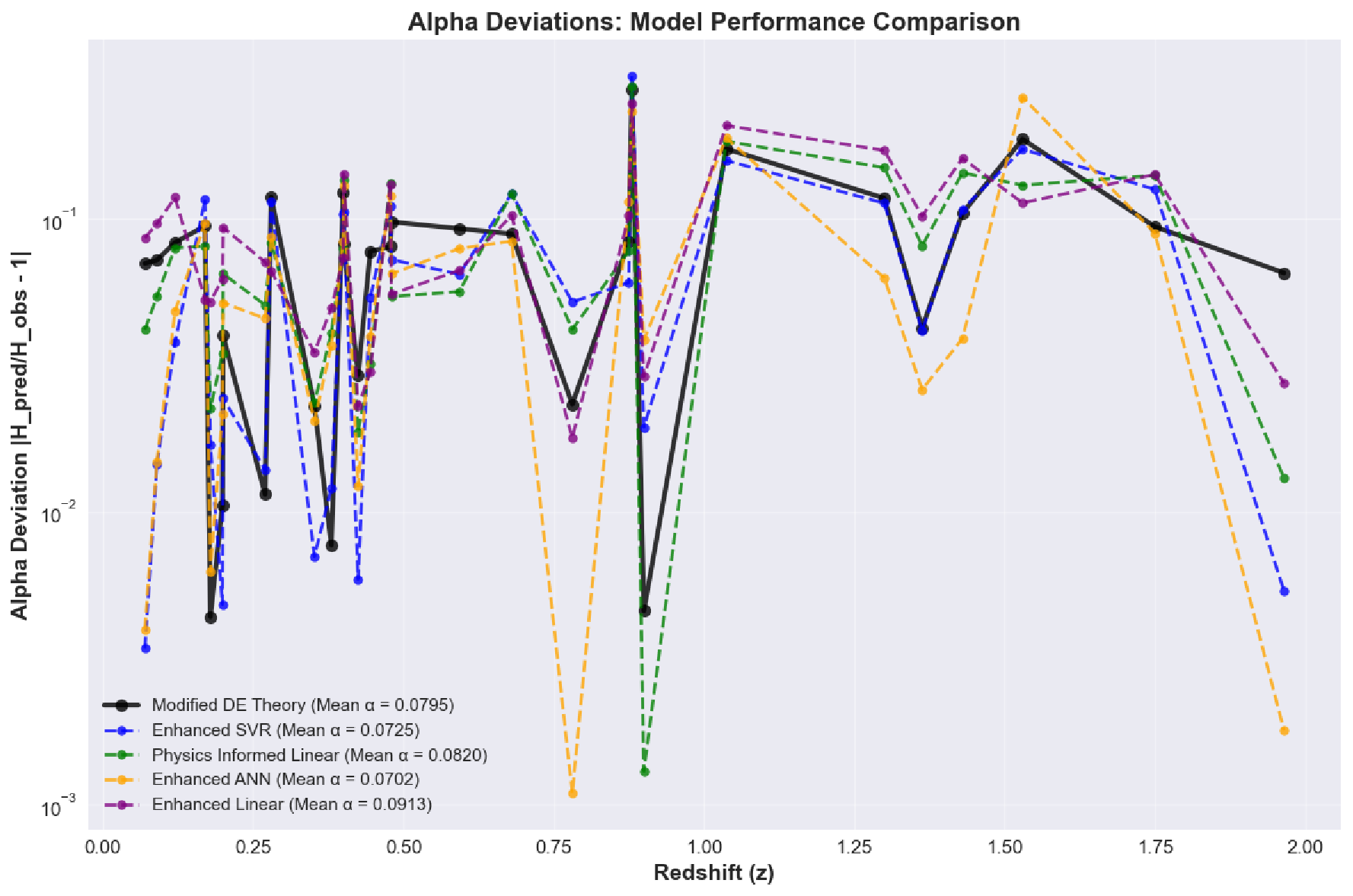}
    \caption{(Left) Machine learning regression fit to observational Hubble parameter $H(z)$ data using our model. The solid black line shows the theoretical prediction, while data points (red circles) correspond to measured $H(z)$ values with $1\sigma$ error bars, the regression fits from ML algorithms are shown using dashed lines, and the close alignment illustrates the model's accuracy in capturing the cosmic expansion history for the Legendre model. (Right) $\chi^2$ goodness of fit for different ML techniques in a bar diagram showing their comparative values for the Legendre model. (Down) Comparison of the absolute relative deviation $|H_{\mathrm{pred}}/H_{\mathrm{obs}} - 1|$ of the Hubble parameter from different predictive models as a function of redshift. The curves represent the theory, Enhanced Support Vector Regression (SVR), Enhanced Artificial Neural Network (ANN), Enhanced Linear Regression, and Physics-Informed Linear Regression. The plot demonstrates that machine learning models, especially Enhanced SVR and ANN, exhibit deviations comparable to the theoretical model, indicating strong predictive performance for the Legendre model.}
    \label{fig:legendre}
\end{figure}

\begin{figure}[h!]
    \centering
    \includegraphics[width=0.48\linewidth]{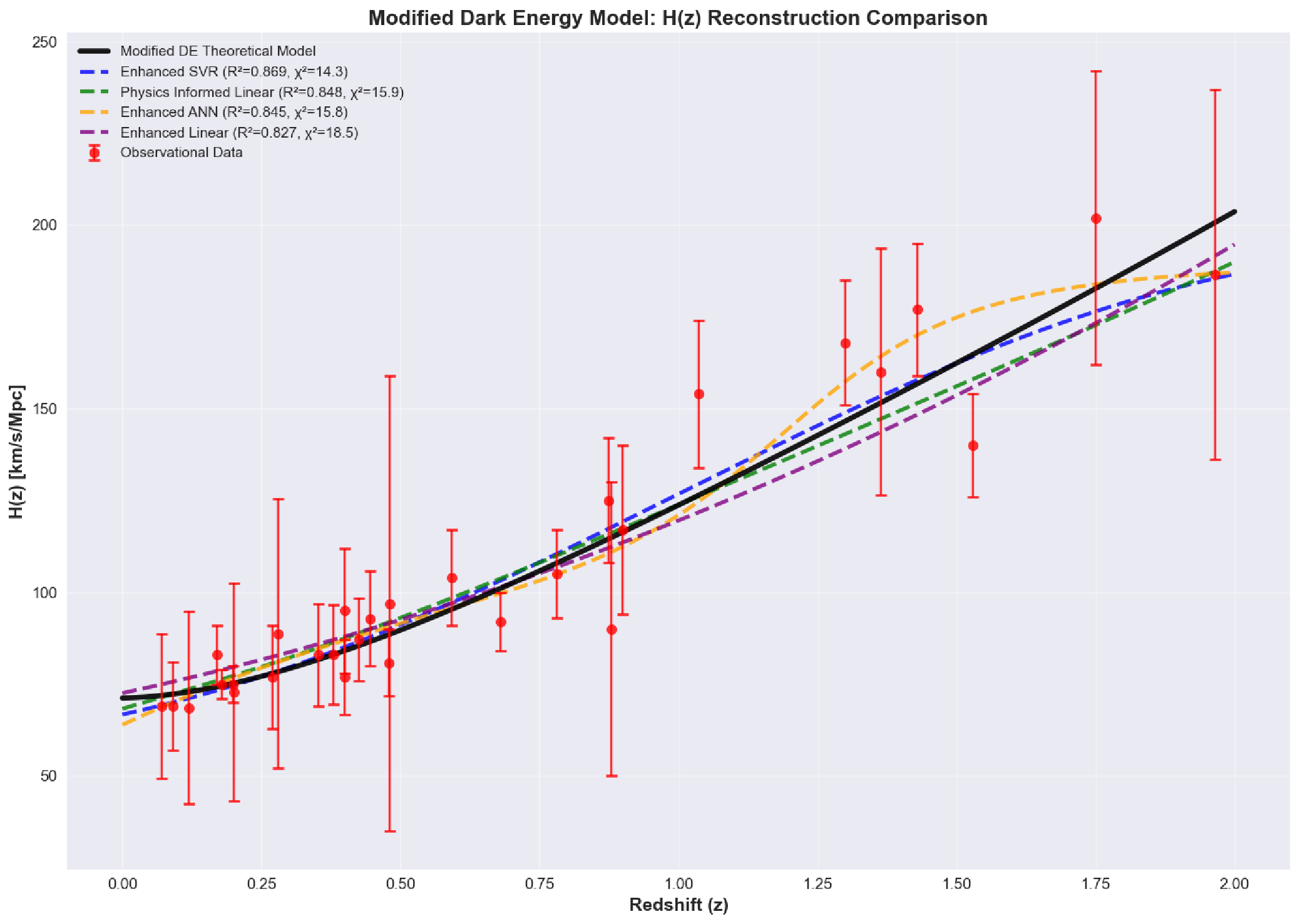}
    \hfill
    \includegraphics[width=0.48\linewidth]{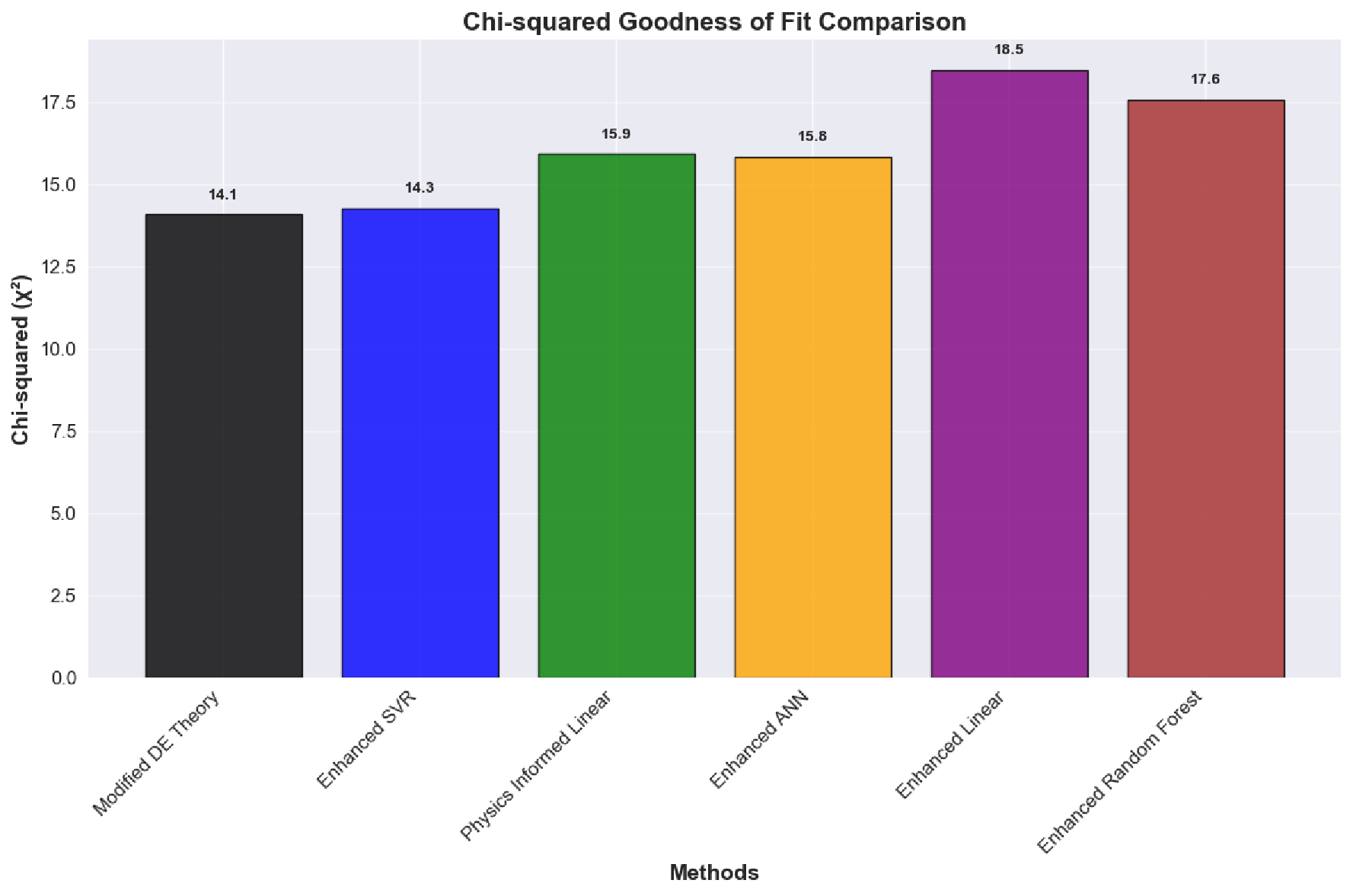}
    \\[0.5cm]
    \includegraphics[width=0.58\linewidth]{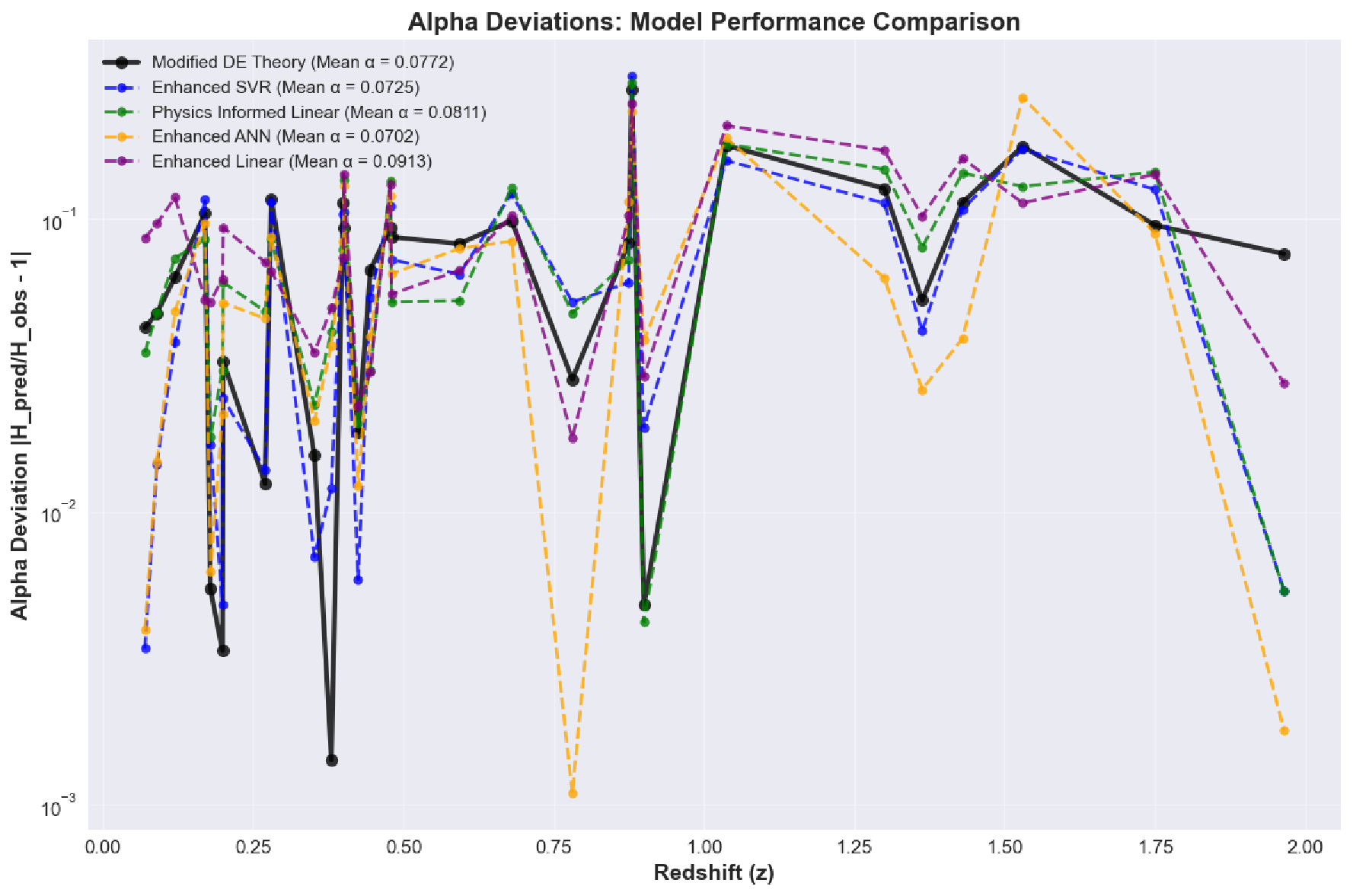}
    \caption{(Left) Machine learning regression fit to observational Hubble parameter $H(z)$ data using our model. The solid black line shows the theoretical prediction, while data points (red circles) correspond to measured $H(z)$ values with $1\sigma$ error bars, the regression fits from ML algorithms are shown using dashed lines and the close alignment illustrates the model's accuracy in capturing the cosmic expansion history for the Laguerre model. (Right) $\chi^2$ goodness of fit for different ML techniques in a bar diagram showing their comparative values for the Laguerre model. (Down) Comparison of the absolute relative deviation $|H_{\mathrm{pred}}/H_{\mathrm{obs}} - 1|$ of the Hubble parameter from different predictive models as a function of redshift. Different curves represent the theory, Enhanced Support Vector Regression (SVR), Enhanced Artificial Neural Network (ANN), Enhanced Linear Regression, and Physics-Informed Linear Regression. The plot demonstrates that machine learning models, especially Enhanced SVR and ANN, exhibit deviations comparable to the  theoretical model, indicating strong predictive performance for the Laguerre model.}
    \label{fig:laguerre}
\end{figure}

\begin{figure}[h!]
    \centering
    \includegraphics[width=0.48\linewidth]{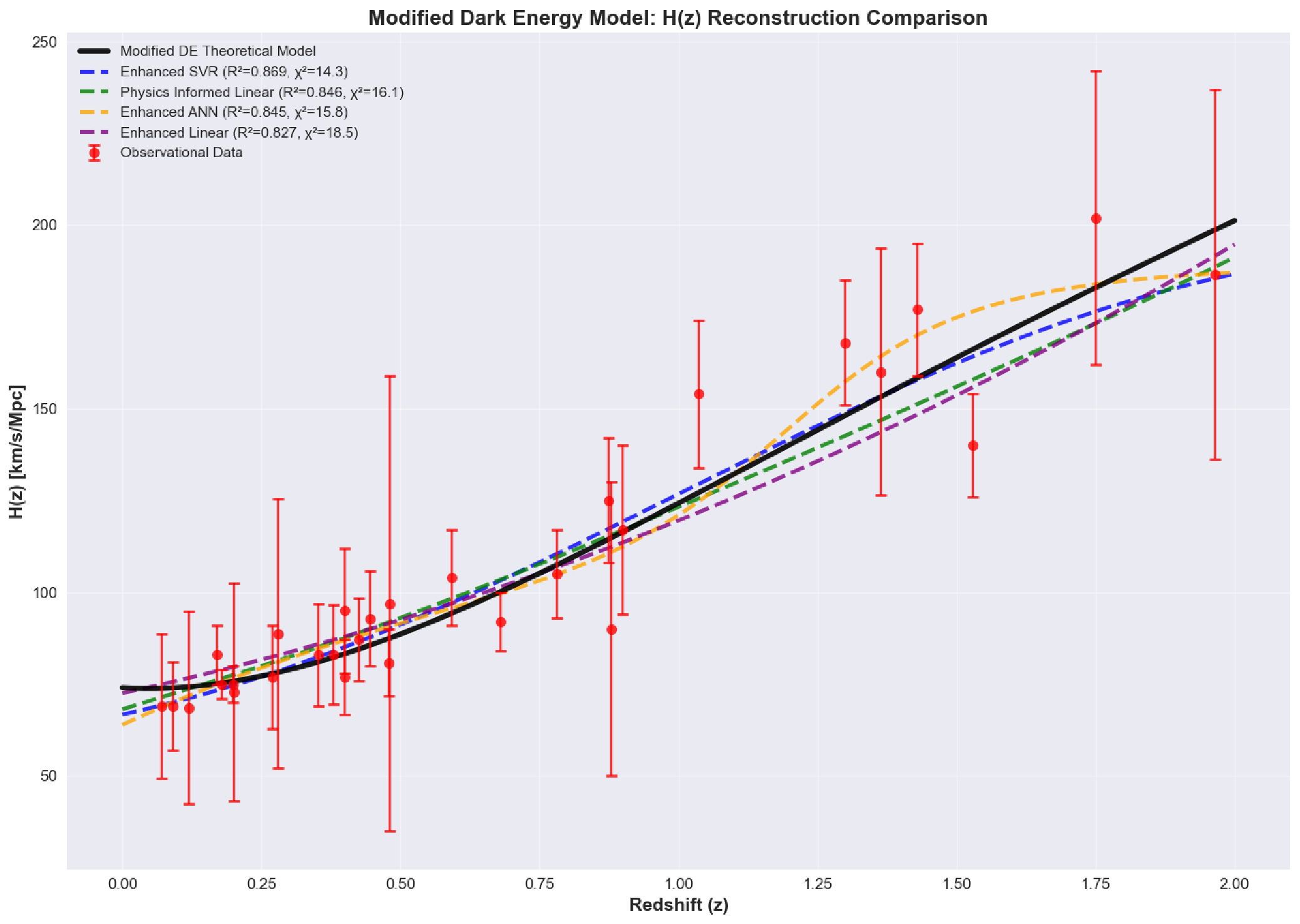}
    \hfill
    \includegraphics[width=0.48\linewidth]{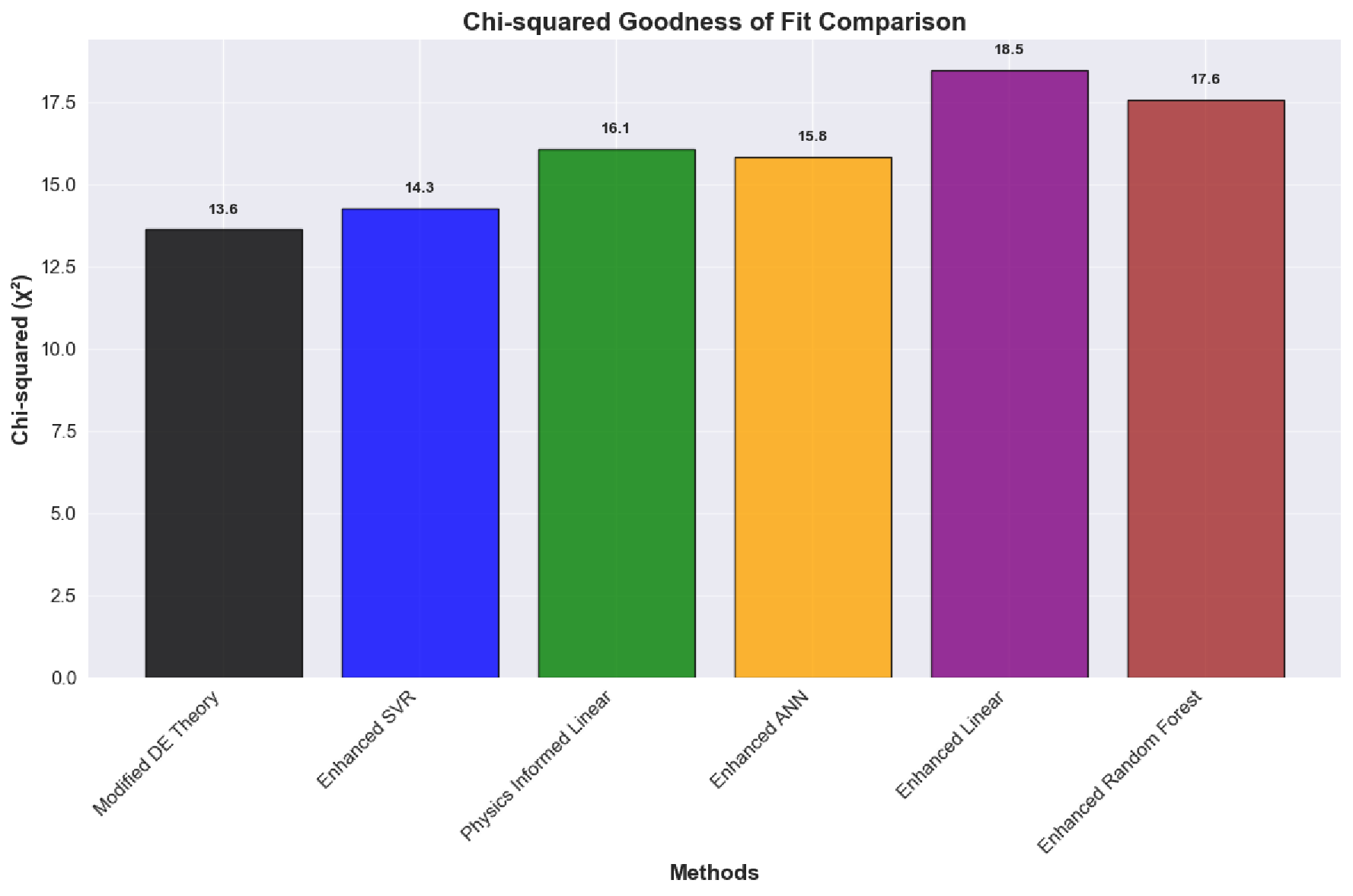}
    \\[0.5cm]
    \includegraphics[width=0.58\linewidth]{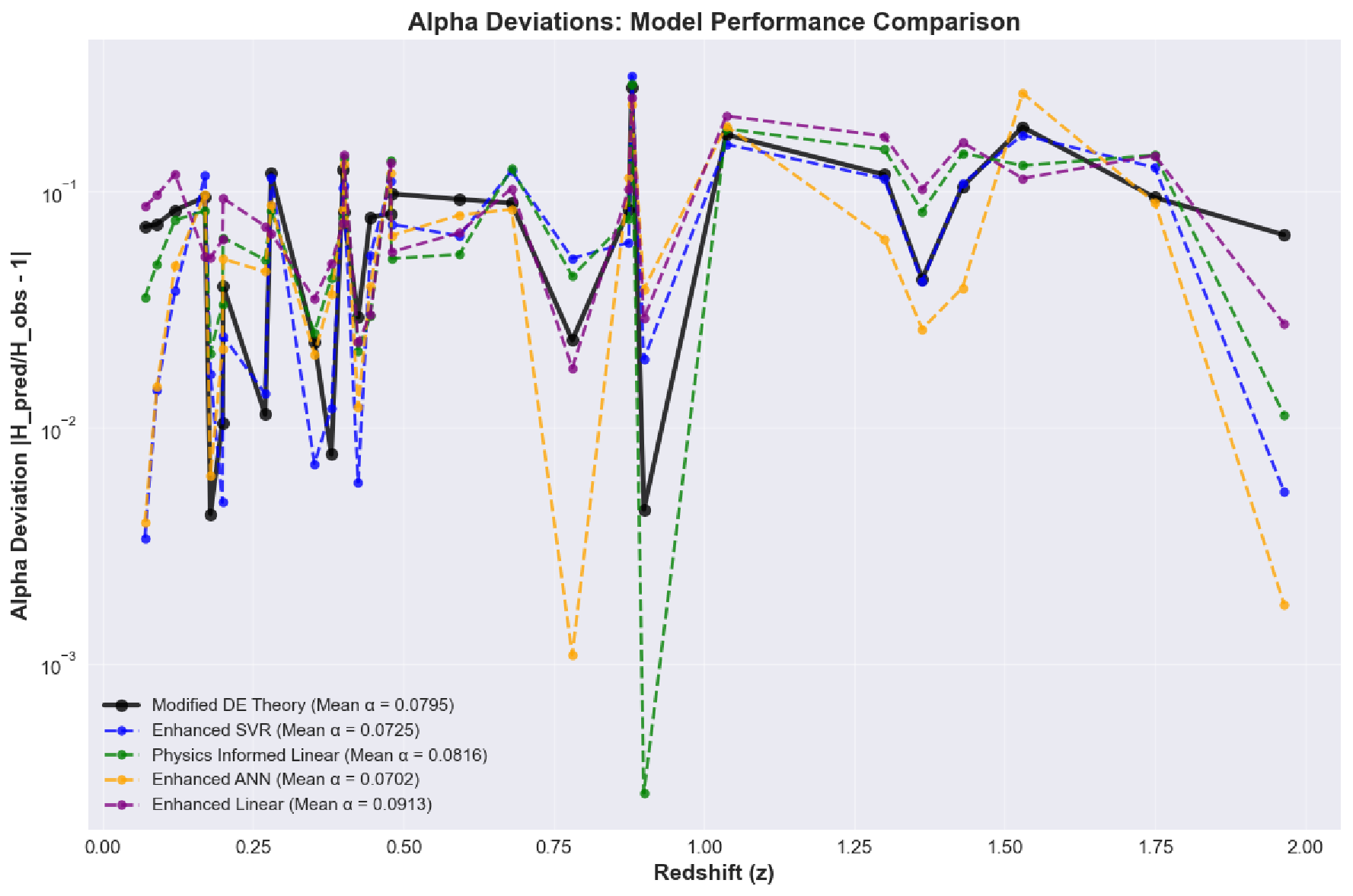}
    \caption{(Left) Machine learning regression fit to observational Hubble parameter $H(z)$ data using our model. The solid black line shows the theoretical prediction, while data points (red circles) correspond to measured $H(z)$ values with $1\sigma$ error bars, the regression fits from ML algorithms are shown using dashed lines, and the close alignment illustrates the model's accuracy in capturing the cosmic expansion history for the Chebyshev model. (Right) $\chi^2$ goodness of fit for different ML techniques in a bar diagram showing their comparative values for the Chebyshev model. (Down) Comparison of the absolute relative deviation $|H_{\mathrm{pred}}/H_{\mathrm{obs}} - 1|$ of the Hubble parameter from different predictive models as a function of redshift. The curves represent the theory, Enhanced Support Vector Regression (SVR), Enhanced Artificial Neural Network (ANN), Enhanced Linear Regression, and Physics-Informed Linear Regression. The plot demonstrates that machine learning models, especially Enhanced SVR and ANN, exhibit deviations comparable to the  theoretical model, indicating strong predictive performance for the Chebyshev model}
    \label{fig:chebyshev}
\end{figure}

\begin{figure}[h!]
    \centering
    \includegraphics[width=0.48\linewidth]{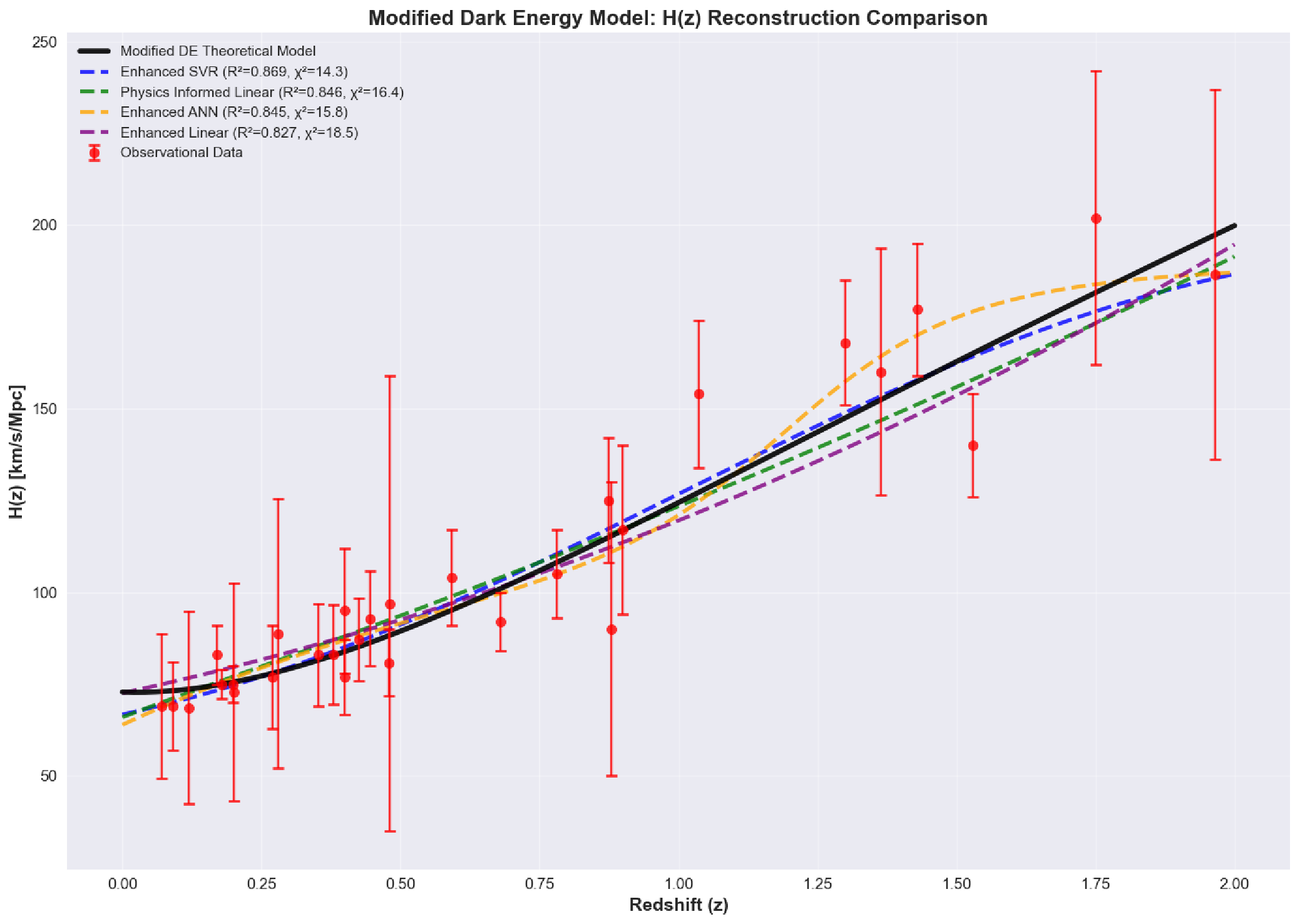}
    \hfill
    \includegraphics[width=0.48\linewidth]{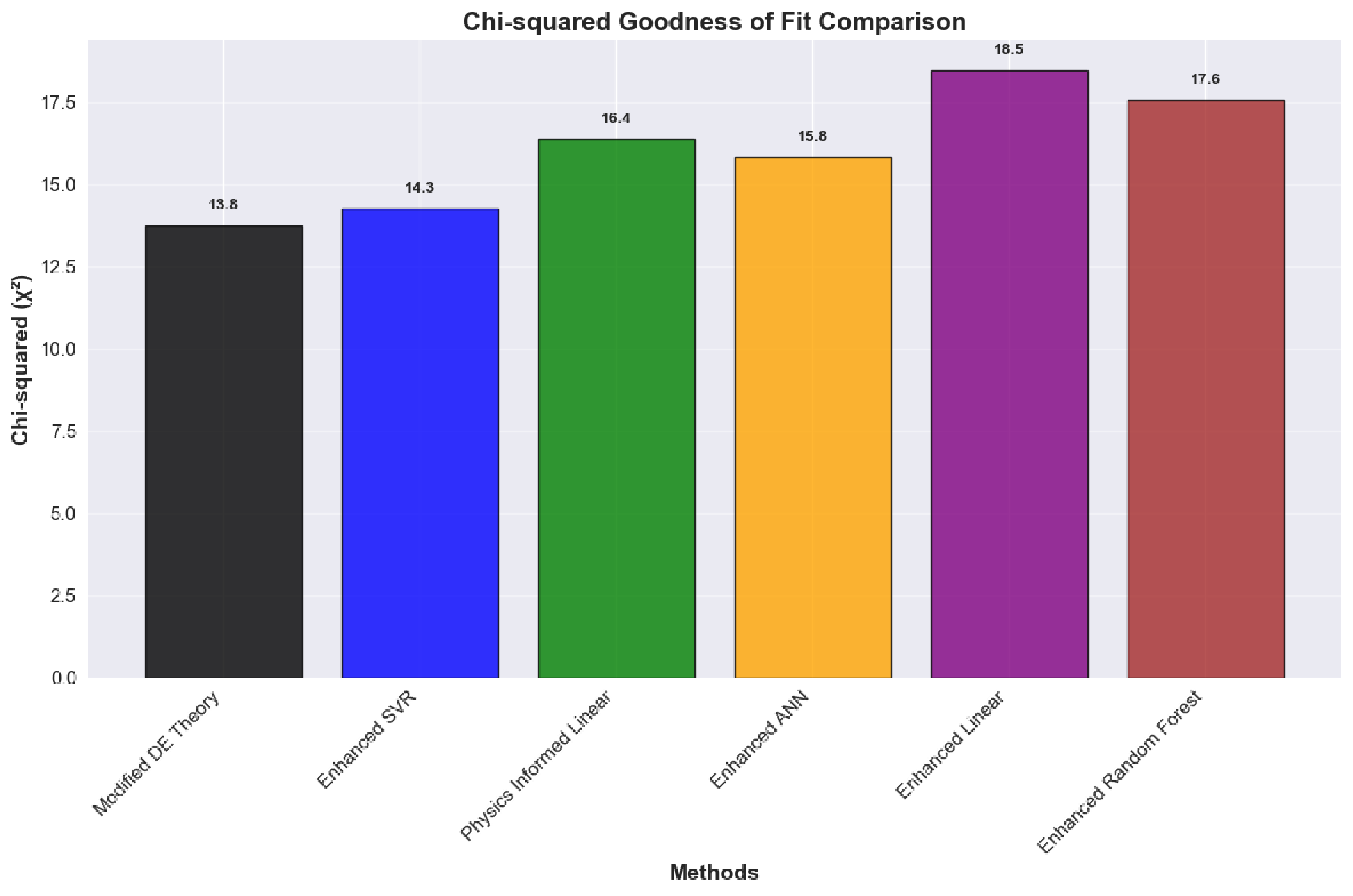}
    \\[0.5cm]
    \includegraphics[width=0.58\linewidth]{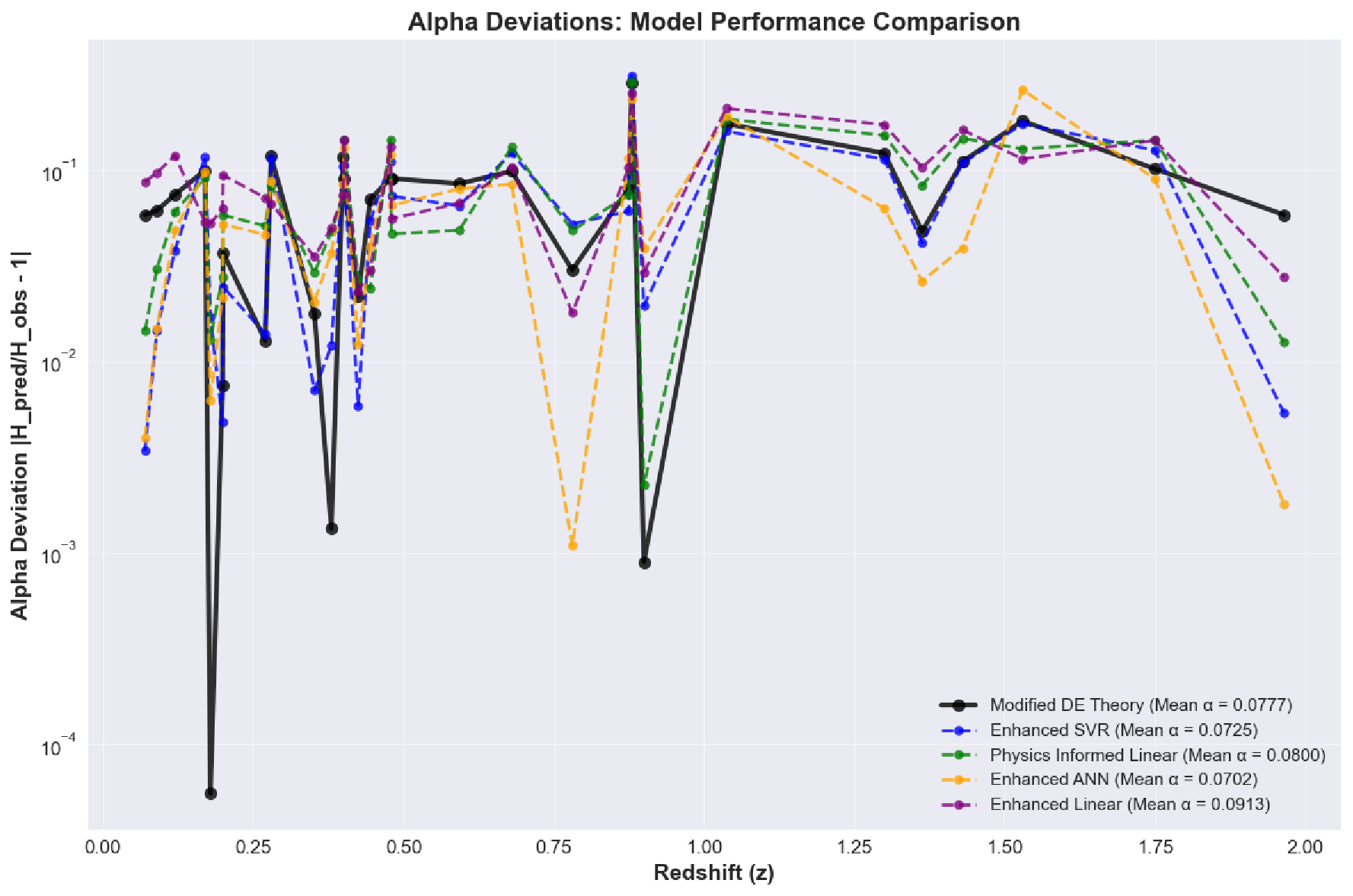}
    \caption{(Left) Machine learning regression fit to observational Hubble parameter $H(z)$ data using our model. The solid black line shows the theoretical prediction, while data points (red circles) correspond to measured $H(z)$ values with $1\sigma$ error bars, the regression fits from ML algorithms are shown using dashed lines, and the close alignment illustrates the model's accuracy in capturing the cosmic expansion history for the Fibonacci model. (Right) $\chi^2$ goodness of fit for different ML techniques in a bar diagram showing their comparative values for the Fibonacci model. (Down) Comparison of the absolute relative deviation $|H_{\mathrm{pred}}/H_{\mathrm{obs}} - 1|$ of the Hubble parameter from different predictive models as a function of redshift. Curves represent the theory, Enhanced Support Vector Regression (SVR), Enhanced Artificial Neural Network (ANN), Enhanced Linear Regression, and Physics-Informed Linear Regression. The plot demonstrates that machine learning models, especially Enhanced SVR and ANN, exhibit deviations comparable to the  theoretical model, indicating strong predictive performance for the Fibonacci model.}
    \label{fig:fibonacci}
\end{figure}

\begin{table}[ht!]
\centering
\caption{Performance Comparison Table (Conventional Basis)}
\scriptsize 
\begin{adjustbox}{max width=\linewidth,center}
\begin{tabular}{lcccccc}
\toprule
\textbf{Algorithm} & \textbf{Train R\textsuperscript{2}} & \textbf{Test R\textsuperscript{2}} & \textbf{Test RMSE} & \textbf{Chi\textsuperscript{2}} & \textbf{Reduced Chi\textsuperscript{2}} & \textbf{Mean $\alpha$ Deviation} \\
\midrule
Enhanced\_SVR              & 0.928013 & 0.869022 & 15.796117 & 14.271957 & 0.492136 & 0.072457 \\
Enhanced\_ANN              & 0.956026 & 0.845141 & 17.175891 & 15.849716 & 0.546542 & 0.070242 \\
Physics\_Informed\_Linear  & 0.912069 & 0.841684 & 17.366594 & 16.622077 & 0.573175 & 0.081319 \\
Enhanced\_Linear           & 0.904353 & 0.826641 & 18.172912 & 18.480340 & 0.637253 & 0.091273 \\
Enhanced\_Random\_Forest   & 0.980640 & 0.786107 & 20.185959 & 17.578790 & 0.606165 & 0.067662 \\
Gradient\_Boosting         & 0.965357 & 0.777222 & 20.600959 & 19.917660 & 0.686816 & 0.081604 \\
Modified\_DE\_Theoretical  & ---      & ---      & 12.547005 & 14.586239 & 0.502974 & 0.077876 \\
\bottomrule
\end{tabular}
\end{adjustbox}
\end{table}

\begin{table}[ht!]
\centering
\caption{Performance Comparison Table (Legendre Basis)}
\scriptsize 
\begin{adjustbox}{max width=\linewidth,center}
\begin{tabular}{lcccccc}
\toprule
\textbf{Algorithm} & \textbf{Train R\textsuperscript{2}} & \textbf{Test R\textsuperscript{2}} & \textbf{Test RMSE} & \textbf{Chi\textsuperscript{2}} & \textbf{Reduced Chi\textsuperscript{2}} & \textbf{Mean $\alpha$ Deviation} \\
\midrule
Enhanced\_SVR              & 0.928013 & 0.869022 & 15.796117 & 14.271957 & 0.492136 & 0.072457 \\
Physics\_Informed\_Linear  & 0.915927 & 0.847133 & 17.065073 & 15.933242 & 0.549422 & 0.081991 \\
Enhanced\_ANN              & 0.956026 & 0.845141 & 17.175891 & 15.849716 & 0.546542 & 0.070242 \\
Enhanced\_Linear           & 0.904353 & 0.826641 & 18.172912 & 18.480340 & 0.637253 & 0.091273 \\
Enhanced\_Random\_Forest   & 0.980640 & 0.786107 & 20.185959 & 17.578790 & 0.606165 & 0.067662 \\
Gradient\_Boosting         & 0.965357 & 0.777222 & 20.600959 & 19.917660 & 0.686816 & 0.081604 \\
Modified\_DE\_Theoretical  & ---      & ---      & \textbf{11.878846} & 13.628197 & \textbf{0.469938} & 0.079474 \\
\bottomrule
\end{tabular}
\end{adjustbox}
\end{table}

\begin{table}[ht!]
\centering
\caption{Performance Comparison Table (Laguerre Basis)}
\scriptsize 
\begin{adjustbox}{max width=\linewidth,center}
\begin{tabular}{lcccccc}
\toprule
\textbf{Algorithm} & \textbf{Train R\textsuperscript{2}} & \textbf{Test R\textsuperscript{2}} & \textbf{Test RMSE} & \textbf{Chi\textsuperscript{2}} & \textbf{Reduced Chi\textsuperscript{2}} & \textbf{Mean $\alpha$ Deviation} \\
\midrule
Enhanced\_SVR              & 0.928013 & 0.869022 & 15.796117 & 14.271957 & 0.492136 & 0.072457 \\
Physics\_Informed\_Linear  & 0.915274 & 0.847581 & 17.040083 & 15.930695 & 0.549334 & 0.081051 \\
Enhanced\_ANN              & 0.956026 & 0.845141 & 17.175891 & 15.849716 & 0.546542 & 0.070242 \\
Enhanced\_Linear           & 0.904353 & 0.826641 & 18.172912 & 18.480340 & 0.637253 & 0.091273 \\
Enhanced\_Random\_Forest   & 0.980640 & 0.786107 & 20.185959 & 17.578790 & 0.606165 & 0.067662 \\
Gradient\_Boosting         & 0.965357 & 0.777222 & 20.600959 & 19.917660 & 0.686816 & 0.081604 \\
Modified\_DE\_Theoretical  & ---      & ---      & 12.036732 & 14.092887 & 0.485962 & 0.077215 \\
\bottomrule
\end{tabular}
\end{adjustbox}
\end{table}

\begin{table}[ht!]
\centering
\caption{Performance Comparison Table (Chebyshev Basis)}
\scriptsize 
\begin{adjustbox}{max width=\linewidth,center}
\begin{tabular}{lcccccc}
\toprule
\textbf{Algorithm} & \textbf{Train R\textsuperscript{2}} & \textbf{Test R\textsuperscript{2}} & \textbf{Test RMSE} & \textbf{Chi\textsuperscript{2}} & \textbf{Reduced Chi\textsuperscript{2}} & \textbf{Mean $\alpha$ Deviation} \\
\midrule
Enhanced\_SVR              & 0.928013 & 0.869022 & 15.796117 & 14.271957 & 0.492136 & 0.072457 \\
Physics\_Informed\_Linear  & 0.915228 & 0.846157 & 17.119490 & 16.085608 & 0.554676 & 0.081624 \\
Enhanced\_ANN              & 0.956026 & 0.845141 & 17.175891 & 15.849716 & 0.546542 & 0.070242 \\
Enhanced\_Linear           & 0.904353 & 0.826641 & 18.172912 & 18.480340 & 0.637253 & 0.091273 \\
Enhanced\_Random\_Forest   & 0.980640 & 0.786107 & 20.185959 & 17.578790 & 0.606165 & 0.067662 \\
Gradient\_Boosting         & 0.965357 & 0.777222 & 20.600959 & 19.917660 & 0.686816 & 0.081604 \\
Modified\_DE\_Theoretical  & ---      & ---      & 11.881022 & 13.632934 & 0.470101 & 0.079499 \\
\bottomrule
\end{tabular}
\end{adjustbox}
\end{table}

\begin{table}[ht!]
\centering
\caption{Performance Comparison Table (Fibonacci Basis)}
\scriptsize 
\begin{adjustbox}{max width=\linewidth,center}
\begin{tabular}{lcccccc}
\toprule
\textbf{Algorithm} & \textbf{Train R\textsuperscript{2}} & \textbf{Test R\textsuperscript{2}} & \textbf{Test RMSE} & \textbf{Chi\textsuperscript{2}} & \textbf{Reduced Chi\textsuperscript{2}} & \textbf{Mean $\alpha$ Deviation} \\
\midrule
Enhanced\_SVR              & 0.928013 & 0.869022 & 15.796117 & 14.271957 & 0.492136 & 0.072457 \\
Physics\_Informed\_Linear  & 0.914785 & 0.846058 & 17.124999 & 16.401544 & 0.565570 & 0.080007 \\
Enhanced\_ANN              & 0.956026 & 0.845141 & 17.175891 & 15.849716 & 0.546542 & 0.070242 \\
Enhanced\_Linear           & 0.904353 & 0.826641 & 18.172912 & 18.480340 & 0.637253 & 0.091273 \\
Enhanced\_Random\_Forest   & 0.980640 & 0.786107 & 20.185959 & 17.578790 & 0.606165 & 0.067662 \\
Gradient\_Boosting         & 0.965357 & 0.777222 & 20.600959 & 19.917660 & 0.686816 & 0.081604 \\
Modified\_DE\_Theoretical  & ---      & ---      & 11.890301 & 13.755043 & 0.474312 & 0.077733 \\
\bottomrule
\end{tabular}
\end{adjustbox}
\end{table}

\subsection{Superior Performance of the Legendre Basis}
Among all tested polynomial bases, the Legendre basis demonstrated the most optimal balance between accuracy, stability, and physical consistency. The MDE theoretical model under the Legendre basis achieved the lowest Test RMSE of 11.8788 and the lowest reduced Chi-square of 0.4699, outperforming all other bases. In comparison, the Conventional basis yielded a Test RMSE of 12.5470 and a reduced Chi-square of 0.5029; the Laguerre basis gave 12.0367 and 0.4859, respectively; the Chebyshev basis reached 11.8810 and 0.4701, respectively; and the Fibonacci basis produced 11.8903 and 0.4743, respectively, all of which are slightly higher than the Legendre results. The superior performance of the Legendre basis can be attributed to its strong orthogonality, uniform weight distribution, and robust numerical stability, which together allow for accurate and physically meaningful approximations. These properties enable the Legendre polynomials to capture complex variations in the model parameters without introducing spurious oscillations or numerical artefacts. Consequently, the Legendre basis provides the most reliable and physically consistent representation of the MDE theoretical framework, and all subsequent analyses, model refinements, and validations in this study are conducted primarily using the Legendre polynomial expansion.

\section{Conclusions}
Even though its theoretical origin has not yet been revealed, phenomenologically parameterizing dark energy offers the chance to advance the characterization of this fundamental component of the universe. The wealth of existing prior knowledge on the subject suggests that parametrizations with two-parameter smooth functions are preferred when adhering to pertinent astrophysical data. Restraint at high redshifts is a second-order criterion, meaning that dark energy shouldn't exhibit a noticeable blueshift or redshift more slowly than matter.

Using these specifications, in this work, we have proposed a smooth, well-behaved redshift function, using which we have developed some redshift parametrizations of the dark energy equation of state. This redshift function is a considerable improvement on the previous polynomial parametrizations, This is basically because it replaces the function $\frac{1}{1+z}$ (which has a future singularity at $z=-1$) with a well-behaved smooth function $\frac{z}{1+z^{2}}$. Then, using some mathematical functions originating from probability, statistics, and solutions of differential equations, we have framed some classes of two-parameter DE EoS. By this, we avoid the high redshift unboundedness while at the same time keeping the function well-behaved throughout the evolution of the universe, i.e. $z ~\epsilon~ [-1,\infty)$. We have used some well-known mathematical functions like the Legendre polynomial, Laguerre polynomial, Chebyshev polynomial, and Fibonacci polynomial to form the parametrized DE EoS models. These functions have wide applications in differential equations, probability theory, statistics, computing, signal processing, mathematical physics, control theory, robotics, cryptography, etc. Since these functions can effectively model various natural processes, there is good motivation to use them in cosmological analysis to check their effectiveness in modelling the dynamics of the universe.

A comprehensive observational data analysis is performed using recent observational data, like Hubble, BAO, and DESI, to constrain the parameter spaces of the proposed models. Confidence contours showing joint and marginalized posterior distribution with different combinations of datasets are generated using a Markov Chain Monte Carlo approach. For all the datasets and their combinations, we have computed the best-fit values and corresponding confidence intervals and presented them in tables. We see that our improved parametrizations enable us to derive more stringent restrictions on the current dark energy EoS and its derivative, which improves performance. Finally, a machine learning analysis is performed using some suitable algorithms like ELR, PILR, ANN, SVR, ERFR and GBR to compare the models. These models were trained using a 20-point training set and evaluated on a 10-point test set. Their performance was assessed through the coefficient of determination ($R^2$), root mean square error (RMSE), chi-squared statistics, reduced chi-squared, and mean absolute deviation in the parameter $\alpha$. The comparative analysis of various machine learning and theoretical models under different polynomial bases—namely, Conventional, Legendre, Laguerre, Chebyshev, and Fibonacci—provided valuable insight into the numerical stability and physical consistency of the proposed modelling framework. Among all the tested polynomial bases, the Legendre basis demonstrated superior performance with the lowest test RMSE ($11.8788$) and reduced chi-square value ($0.4699$) under the Modified Differential Evolution (MDE) theoretical model, indicating exceptional physical accuracy and numerical stability. The Enhanced Support Vector Regression (SVR) and Artificial Neural Network (ANN) models also showed strong predictive power with high $R^2$ values ($\approx 0.87$) and low RMSE, validating their ability to replicate the theoretical trends. The integration of observational constraints, polynomial parametrizations, and physics-informed machine learning techniques thus establishes a powerful hybrid framework that enhances both the precision and interpretability of cosmological model reconstruction, offering a consistent bridge between data-driven and theory-based approaches to dark energy evolution. As a future project, we would like to continue exploring these models with other datasets and other refined data analysis methods to test the reliability of the results obtained in this work. Studying the cosmological evolution of these models can be a very important future project with these models.

\section*{Acknowledgments}
P.R. acknowledges the Inter-University Centre for Astronomy and Astrophysics (IUCAA), Pune, India, for granting a visiting associateship. All the authors acknowledge the hospitality provided by IUCAA during a visit, when a majority portion of this work was done.

\section*{Data Availability Statement}

All the data analyzed in this study have been properly referenced in the paper.

\section*{Conflict of Interest}

There are no conflicts of interest.

\section*{Funding Statement}

There is no funding to report for this article.

\bibliographystyle{ieeetr} 


\end{document}